\documentclass[useAMS,usenatbib,a4]{mn2e}
\usepackage{times}
\usepackage{abbrev}

\usepackage{amssymb}
\usepackage{graphicx}
\usepackage{path}

\usepackage[usenames]{color}

\title[The VeLLO L1148--IRS]{Confirmation of the VeLLO L1148--IRS: Star
  Formation at very low (Column) Density}

\author[J.\ Kauffmann et al.]{
  J.\ Kauffmann$^1$\thanks{E-mail: jens.kauffmann@jpl.nasa.gov},
  F.\ Bertoldi$^2$,
  T.L.\ Bourke$^3$,
  P.C.\ Myers$^3$,
  C.W.\ Lee$^4$,
  T.L.\ Huard$^5$\\
  $^1$NPP Fellow, Jet Propulsion Laboratory, California Institute of
  Technology, 4800 Oak Grove Drive, Pasadena, CA 91109, USA\\
  $^2$Argelander Institut f\"ur Astronomie, Universit\"at Bonn, Auf dem
  H\"ugel 71, 53121 Bonn, Germany\\
  $^3$Harvard--Smithsonian Center for Astrophysics, 60 Garden Street,
  Cambridge, MA 02138, USA\\
  $^4$Korea Astronomy and Space Science Institute, 61--1 Hwaam--dong,
  Yusung--gu, Daejeon 305--348, Korea\\
  $^5$Department of Astronomy, University of Maryland,
  College Park, MD 20742, USA
}

\begin{document}

\date{Accepted XXX. Received YYY; in original form ZZZ}

\pagerange{\pageref{firstpage}--\pageref{lastpage}} \pubyear{2010}

\maketitle
\label{firstpage}

\begin{abstract}
  We report the detection of a compact ($\sim{}5\arcsec$; about
  $1800~\rm{}AU$ projected size) CO outflow from L1148--IRS. This
  confirms that this Spitzer source is physically associated with the
  nearby ($\approx{}325~\rm{}pc$) L1148 dense core. Radiative transfer
  modeling suggests an internal luminosity of
  $0.08~{\rm{}to}~0.13\,L_{\sun}$. This validates L1148--IRS as a Very
  Low Luminosity Object (VeLLO; $L\le{}0.1\,L_{\sun}$). The L1148
  dense core has unusually low densities and column densities for a
  star--forming core. It is difficult to understand how L1148--IRS might
  have formed under these conditions. Independent of the exact final
  mass of this VeLLO (which is likely $<0.24\,M_{\sun}$), L1148--IRS
  and similar VeLLOs might hold some clues about the isolated
  formation of brown dwarfs.
\end{abstract}

\begin{keywords}
stars: formation;
stars: low-mass, brown dwarfs;
ISM: globules;
ISM: individual: L1148;
ISM: jets and outflows;
infrared: stars
\end{keywords}

\section{Introduction}
The sensitivity of the Spitzer Space Telescope allows for searches and
studies of young stellar objects that are much less luminous than
those studied before. While surveys of star--forming regions using data
from the Infrared Astronomical Satellite (IRAS) were sensitive to
sources with luminosities
$\gtrsim{}0.1\,L_{\sun}\,(d/{140~\rm{}pc})^2$ (where $d$ is the
distance; \citealt{myers1987:nir_iras}), Spitzer can detect objects
with luminosities as low as
$4\times{}10^{-3}\,L_{\sun}\,(d{}/{140~\rm{}pc})^2$, i.e.\ has a
sentitivity better by a factor 25 \citep{dunham2008:vellos}. Spitzer
has imaged a large number of isolated dense cores thought to be
starless. About 60 of them were observed as part of the Spitzer Legacy
Project ``From Molecular Cores to Planet Forming Disks'', or c2d
\citep{evans2003:c2d}.

These observations led to a series of discoveries of Very Low
Luminosity Objects (VeLLOs; for definitions see
\citealt{kauffmann2005:vellos}, \citealt{difrancesco2006:ppv},
\citealt{dunham2008:vellos}), infrared stars of luminosity
$\le{}0.1\,L_{\sun}$ which are embedded in dense cores. VeLLO colors
and envelope properties are consistent with those of class 0 or class
I protostars \citep{lada1987:ass_to_protostars, andre1993:class0},
young stars in their main accretion phase that are deeply embedded in
the dense core from which they form.

\citet{young2004:l1014} presented the first Spitzer--identified VeLLO,
L1014--IRS (also see: \citealt{crapsi2005:l1014, huard2006:l1014,
  bourke2005:l1014-outflow}). Further detailed studies were presented
for L1521F--IRS \citep{bourke2006:l1521f} and L328--IRS
\citep{lee2009:l328}. These papers established the initial VeLLO
  sample and the first overviews of their properties. Other
well--studied VeLLO candidates are Cha--MMS1
\citep{belloche2006:cha-mm1} and L673--7--IRS
\citep{dunham2010:l673-7}. See Table \ref{tab:vello-properties} for an
up--to--date list. The combined c2d survey data for cores and clouds
reveals 15 candidate VeLLOs \citep{dunham2008:vellos}. Not all of them
are well studied, though. The first VeLLO ever discovered was
IRAM04191. It was identified by \citet{andre1999:iram04191} on the
basis of millimeter radio observations (see
\citealt{dunham2006:iram_04191} for Spitzer results).\medskip

\noindent{}The evolutionary status and fate of VeLLOs is presently
unclear and likely to be diverse. There is probably general consent
that VeLLOs have a present--day mass $<0.1\,M_{\sun}$
(\citealt{young2004:l1014}; also see Section
\ref{sec:stage-mass}). However, some VeLLOs drive prominent outflows,
and thus seem to accrete significantly. These objects have been
interpreted as very young ($\sim{}10^4~\rm{}yr$) protostars of stellar
final mass (e.g., IRAM04191, which has a prominent outflow;
\citealt{andre1999:iram04191, belloche2002:iram04191}). In contrast,
VeLLOs devoid of such outflows might not accrete at a high rate. Some were
proposed to be proto brown dwarfs forming in isolation (i.e., not as
low--mass components of clusters and binaries; e.g., L1014--IRS, which
has a $\sim{}3\arcsec$ [$\sim{}750~\rm AU$] outflow only detected with
interferometers; \citealt{young2004:l1014, crapsi2005:l1014,
  bourke2005:l1014-outflow, huard2006:l1014}). This concept has been
bolstered by recent discoveries of binary brown dwarfs which are too
loosely bound to have been expelled by multiple stellar systems (e.g.,
\citealt{radigan2009:wide-bd} and \citealt{luhman2009:fu-tau}; also
see \path|http://www.vlmbinaries.org/|).

\begin{figure}
\includegraphics[height=\linewidth,angle=-90]{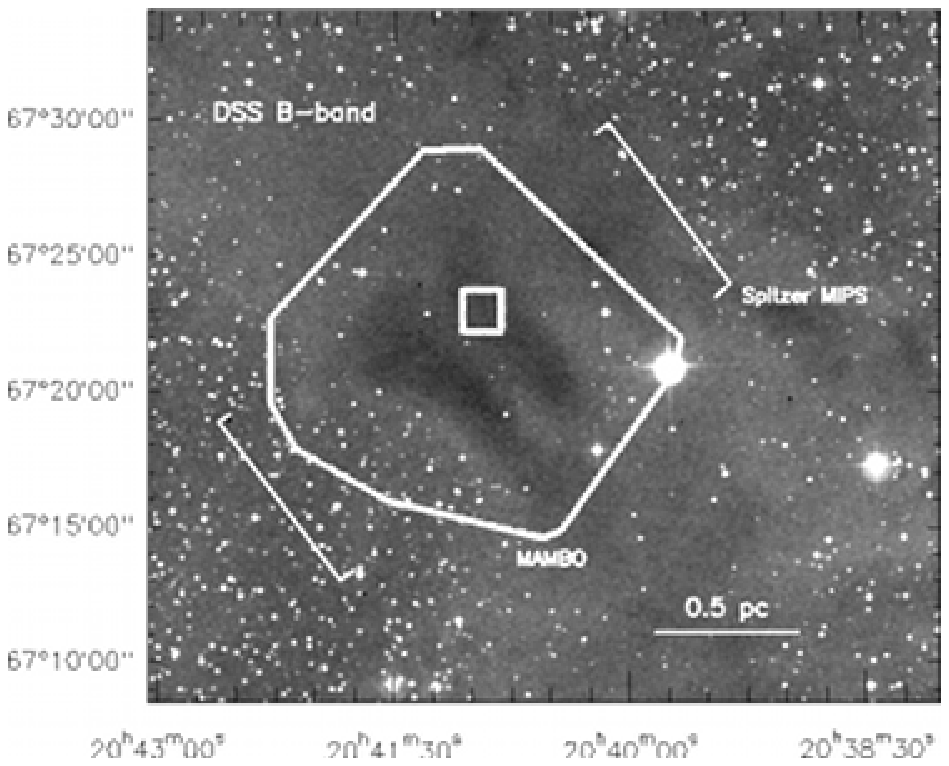}
\vspace{0.1cm}\\
\includegraphics[width=0.448\linewidth,angle=-90]{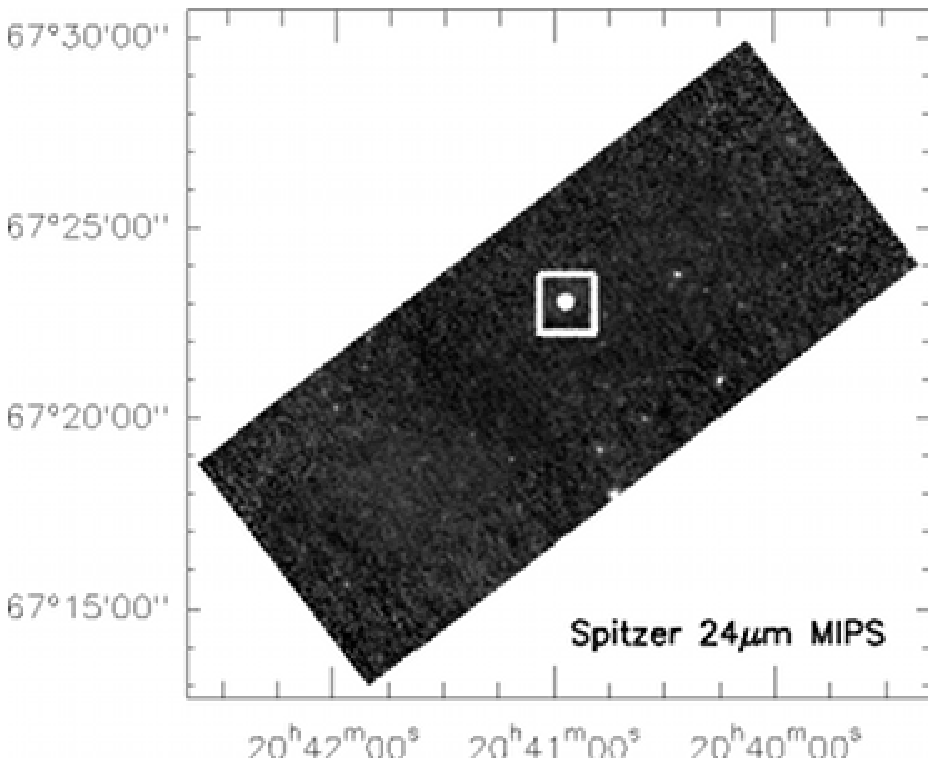}
\includegraphics[width=0.448\linewidth,angle=-90]{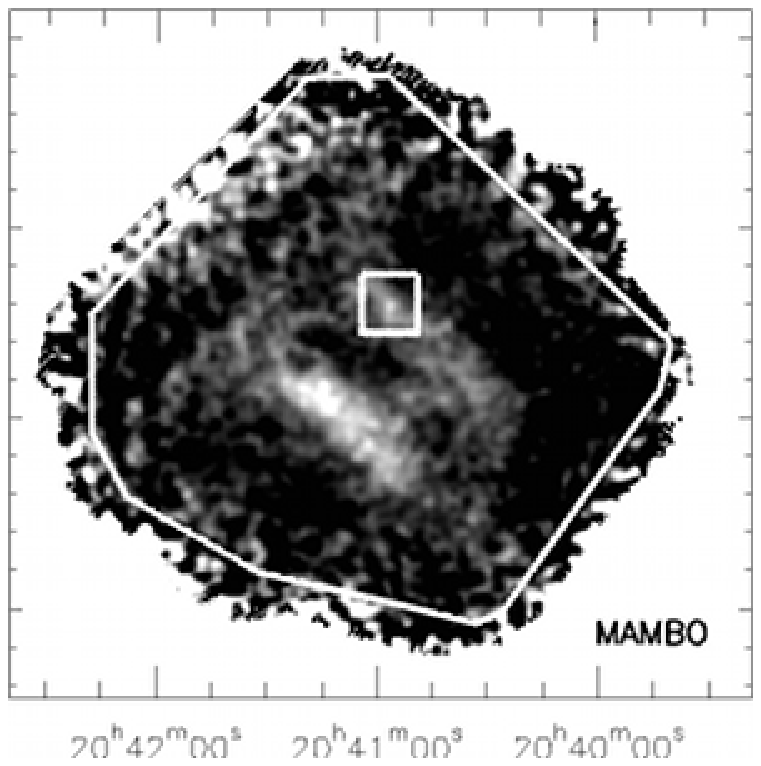}
\caption{The discovery of L1148--IRS. The L1148 region contains two
  dark dust features evident at optical wavelength (top panel). These
  are also detected in maps of dust continuum emission (bottom right
  panel). At $24~\rm{}\mu{}m$ wavelength, the source L1148--IRS (framed
  by a box in all panels) manifests as an unusually bright point
  source (bottom left panel). It resides near a relatively faint
  ($9~\rm{}mJy$ per $11\arcsec$ beam at $1.2~\rm{}mm$ wavelength after
  smoothing to $20\arcsec$ resolution) dust continuum emission peak
  (bottom right panel). Solid and dotted lines indicate the relative
  orientation of the different maps. See Figure
  \ref{fig:layout-observations} for further information. Coordinates
  are given for J2000.\label{fig:discovery}}
\end{figure}

The second Spitzer VeLLO to be discovered --- the source discussed in
the present paper, L1148--IRS (\citealt{kauffmann2005:vellos}; first
noticed by \citealt{kun1998:cep_flare}) --- always maintained a
peculiar status. As shown in Figure \ref{fig:discovery}, this object
was discovered as an unusually bright Spitzer MIPS source (compared to
the field stars), SSTc2d J204056.66+672304.9, which is closely
associated with a $1.2~\rm{}mm$ wavelength dust emission peak revealed
by the c2d MAMBO survey \citep{kauffmann2008:mambo-spitzer}. It
differs from other VeLLOs in that no other such object is as
inconspicuous in dust emission \citep{kauffmann2005:vellos,
  kauffmann2008:mambo-spitzer, wu2007:c2d_sharc-ii}. This implies that
L1148--IRS resides in a natal dense core of unusually low mass and
average density ($\sim{}0.15\,M_{\sun}$, few $10^4~\rm{}cm^{-3}$;
\citealt{kauffmann2008:mambo-spitzer}).

Unfortunately, initial single--dish searches by
\citet{kauffmann2005:vellos} did not reveal any molecular
outflow. Thus, \citet{kauffmann2005:vellos} could not entirely rule
out that L1148--IRS is just a galaxy (probability $\sim{}10^{-4}$,
using $24~\rm{}\mu{}m$ source counts by
\citealt{papovich2004:galaxy_counts}) or background protostar
accidentally aligned with the L1148 dense core. Formally, this means
that L1148--IRS does not meet all of the \citet{dunham2008:vellos}
criteria for confirmed VeLLOs. Confirmation would require an outflow
detection towards L1148--IRS at the systemic velocity of the dense
core. This would prove that L1148--IRS (\textit{i}) is a YSO which is
(\textit{ii}) physically associated with the L1148 region. Here, we
present an outflow meeting these criteria.\medskip

\noindent{}Our study is organized as follows. Section
\ref{sec:observations} introduces the data on which our study is
based. We discuss the discovery and characterization of the outflow in
Section \ref{sec:outflow}. The physical properties of the embedded YSO
are derived in Section \ref{sec:sed}. Information on various VeLLO
  natal cores are summarized in Section \ref{sec:catalogue}. This
  provides a catalogue against which L1148 can be compared. In
Section \ref{sec:discussion} we turn to a discussion of L1148--IRS in
the context of other VeLLOs. Section \ref{sec:summary} summarizes our
study.

In order to improve the readability of this paper, some complementary
material is removed to two appendices. Appendix \ref{app:photometry}
provides details on the photometric data used in this study. Appendix
\ref{app-sec:dense-core} comprehensively presents data for which a
cursory analysis was executed in previous studies (i.e.,
\citealt{kauffmann2005:vellos}), and describes new data indirectly
supporting the outflow analysis executed here.

\section{Data \& Observations}
\label{sec:observations}

\subsection{Distance \& Previous Studies\label{sec:previous-work}}
\citet{straizys1992:distances} derive a distance of
$(325\pm{}25)~\rm{}pc$ for the L1148 region. This is smaller than the
$450~\rm{}pc$ estimated by \citet{kun1998:cep_flare}. We adopt the
result by \citet{straizys1992:distances}, since they focus on a
$\le{}1\degr\times{}1\degr$ field centered on L1148, while
\citet{kun1998:cep_flare} study an $\approx{}2{}\degr\times{}10\degr$
region in which L1148 lies only at the edge. This is consistent with a
preliminary 2MASS--based distance estimate of $(301\pm{}55)~\rm{}pc$
(G.\ Maheswar \& C.W.\ Lee, in prep.; based on
\citealt{maheswar2010:distances}).

The L1148 dense core was included in the \citet{lee1999:contr_survey,
  lee2001:contr_survey, lee2004:contr_survey} surveys for inward
motions. No signs of such motions were found, but the observed lines
were weak and the position of L1148--IRS was not probed.
\citet{kirk2005:scuba_survey} observed L1148 with SCUBA at
$450~\rm{}\mu{}m$ and $850~\rm{}\mu{}m$ wavelength, but their small
target field ($\approx 2 \arcmin$ diameter) did not include the VeLLO
candidate considered here. The L1148--IRS source was then presented by
\citet{kauffmann2005:vellos}. \citet{wu2007:c2d_sharc-ii} did observe
this source with SHARC--II at $350~\rm{}\mu{}m$ wavelength, but failed
to detect it. These data, combined with the characteristics of the
Spitzer--detected source, led \citet{dunham2008:vellos} to the
conclusion that L1148--IRS ``has a high likelihood of being an embedded
low-luminosity protostar'' (their `group 3'). The source is also
included in the $1.8~\rm{}cm$ radio continuum surveys by
\citet{scaife2010:ami-protostars}. L1148--IRS is detected at a
signal--to--noise ratio $\sim{}3$. As shown by
\citeauthor{scaife2010:ami-protostars}, and repeated in Table
\ref{tab:vello-properties}, the implied intrinsic luminosity at
$1.8~\rm{}cm$ wavelength is comparable to what is found for other
VeLLOs (i.e.\ $F_{1.8\rm{}cm}d^2$, the product of observed flux
density and distance squared).

Most recently, \citet{kirk2009:cepheus-flare} used data from the
Spitzer Gould Belt Legacy survey to search the entire L1148 region for
young stellar objects (YSOs). Just a single candidate YSO, i.e.\ our
L1148--IRS, is found.  This dearth of YSOs is confirmed by
\citet{dunham2008:vellos}, though they find one further candidate with
rather low YSO probability (i.e., in their `group 6'). Like
neighboring clouds, L1148 does thus form stars only at a very small
rate or efficiency \citep{kirk2009:cepheus-flare}.
\citet{nutter2009:cepheus-flare} took Akari data at wavelength of
65 to $160 \micron$ towards a $\sim 0\fdg{}5 \times 0\fdg{}5$ field
containing L1148--IRS. They detect and confirm L1148--IRS; they do,
however, focus on the extended matter in this region and uncover
evidence for external heating of the clouds.\medskip

\noindent{}Detailed assessments of the properties of the natal dense
core of VeLLOs were first presented by
\citet{kauffmann2005:vellos}. Based on maps in lines of CCS (from the
the Effelsberg 100m--telescope) and $\rm{}N_2H^+$ (from the IRAM
30m--telescope), the dense core has line widths at half intensity
$\lesssim{}0.25~\rm{}km\,{}s^{-1}$. Non--thermal motions are thus
sub--sonic, since $\rm{}H_2$ at $10~\rm{}K$ temperature has a line
width of $0.44~\rm{}km\,{}s^{-1}$. Towards L1148--IRS, $\rm{}N_2H^+$
lines yield a systemic velocity of
$2.60~\rm{}km\,{}s^{-1}$. Significant velocity gradients and jumps, as
well as multiple velocity components in $\rm{}C^{18}O$, are observed
near the YSO. The letter by \citet{kauffmann2005:vellos} unfortunately
lacked the space to present the data in detail. These are therefore
for the first time provided in Appendix \ref{app-sec:dense-core}.

Based data from the IRAM 30m--telescope, \citet{kauffmann2005:vellos}
found no indications for significant chemical evolution. Specifically,
CO is depleted by a factor less than 10, and $\rm{}N_2D^+$ is not
detected. See Appendix \ref{app-sec:abundances} for more
details. However, there is evidence for contraction motions within
L1148: line asymmetries indicative of such motions are observed
towards all three positions highlighted in Fig.\
\ref{fig:layout-observations}. These new data are presented in
Appendix \ref{app-sec:contraction}.

The \citet{kauffmann2008:mambo-spitzer} dust emission maps (obtained
using MAMBO at the IRAM 30m--telescope) reveal an unusually small mass
reservoir for a star--forming core. Division of the peak $\rm{}H_2$
column density of $6\times{}10^{21}~\rm{}cm^{-2}$ by the filament
minor and major axis of the parental core
($1\arcmin{}\times{}5\arcmin$ angular size; Figs.\ \ref{fig:discovery}
and \ref{fig:layout-observations}, Appendix
\ref{app-sec:mass-distribution}) suggest mean $\rm{}H_2$ densities of
$(0.4$--$2.0)\times{}10^4~\rm{}cm^{-3}$. Comparison with model density
distributions with central density plateau and
${\rm{}d}\ln(\varrho)/{\rm{}d}\ln(r)=-2$ for large radii (Fig.\ 5 of
\citealt{kauffmann2008:mambo-spitzer}) give a central density
$<3\times{}10^4~\rm{}cm^{-3}$. For a source--centered aperture of
$4200~\rm{}AU$ radius, a mass of only $(0.14\pm{}0.02)\,M_{\sun}$ is
derived\footnote{The aperture radius of $4200~\rm{}AU$ was
    defined by \citet{motte2001:protostars}. In Taurus, this size
    conveniently corresponds to a radius of $30\arcsec$. Also,
    \citeauthor{motte2001:protostars} argue that this radius roughly
    contains the mass that can be accreted in about $10^5~\rm{}yr$ of
    free--fall collapse of an initially hydrostatic sphere of
    $\sim{}0.2~\rm{}km\,s^{-1}$ sound speed. However, except for these
    general considerations, the radius is somewhat arbitrary. We adopt
    this size to compare with previous observations.}. Division of this aperture mass by the aperture
area gives an aperture--averaged column density of
$\langle{}N({\rm{}H_2})\rangle_{4200~\rm{}AU}=4.8\times{}10^{21}~\rm{}cm^{-2}$.

These values are unusually low, even for starless cores
\citep{kauffmann2008:mambo-spitzer}. L1148--IRS thus resides in a natal
core of unusually low density and column density. In this respect,
L1148 is similar to the VeLLO core L1014, which was deemed to be
``starless'' previous to Spitzer observations. This makes it so
interesting.

The mass concentration appears to be centrally peaked: as explained in
Appendix \ref{app-sec:mass-distribution}, cuts through the emission
suggest density profiles
${\rm{}d}\ln(\varrho)/{\rm{}d}\ln(r)\sim{}-2$. L1148--IRS actually
resides $7\farcs{}6$ (i.e., $2500~\rm{}AU$) offset from the dust
emission peak. While this offset is potentially interesting,
\citet{kauffmann2008:mambo-spitzer} use Monte--Carlo
  experiments to show that this dislocation is only marginally
significant given the uncertainties of their data: the beam of
  $11\arcsec$ size, the low signal--to--noise ratio, possible
  anomalous refraction, and potential pointing errors of order
  $3\arcsec$ have to be taken into account. In combination, these
  effects can lead to apparent offsets between the VeLLO and dust
  emission peak position, even if such offsets do not exist.

\subsection{Spitzer Space Telescope}
L1148 was observed by the Spitzer Space Telescope with the Infrared
Array Camera (IRAC; AORKEYs 5158144) and the Multiband Imaging
Photometer for Spitzer (MIPS; AORKEY 9440768) as part of the c2d
Legacy program (\citealt{evans2003:c2d}, PID 139). Further Spitzer
observations (AORKEYs 14608128, 14615808, 18160384) were acquired by
the ``cores2deeper'' (deep IRAC and MIPS integrations; PID 20386) and
``Going Long'' (first MIPS $160~\rm{}\mu{}m$ images of the region; PID
30384) programs. Initial versions of these maps were used in
  \citet{kauffmann2005:vellos}. The images presented here are derived
  from new mosaics combining all of the aforementioned data.
Independent imaging by the Spitzer Gould Belt Legacy Survey
(Allen et al., PID 30574) is discussed by
\citet{kirk2009:cepheus-flare}. Spectra of L1148--IRS have been
acquired too (PID 30563; AORKEY 18401280), but are not discussed
here. Details of the observations, post--processing and source
extraction are presented by \citet{evans2007:c2d-delivery}.

\subsection{IRAM 30m--telescope}
\begin{figure}
\includegraphics[width=\linewidth]{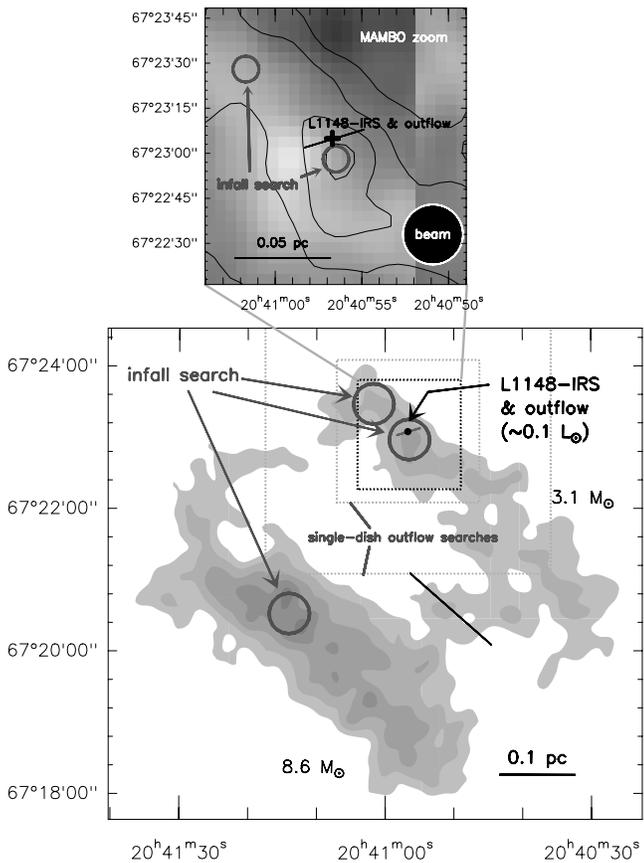}
\caption{Overview of observing strategy and results. As shown by the
  grey shading in the bottom main panel, the region contains two
  parallel filaments detected in dust continuum emission. They have
  masses of $8.6\,M_{\sun}$ and $3.1\,M_{\sun}$, respectively (a black
  solid line marks the adopted boundary between the filaments). The
  position of L1148--IRS and its outflow are indicated by dots and
  lines, respectively. Boxes centered on L1148--IRS indicate the
  regions searched for outflows. Circles mark positions probed for
  dense core contraction motions (Section \ref{sec:previous-work},
  Appendix \ref{app-sec:contraction}). The top zoom panel presents
  details of the MAMBO dust emission map in greyscale and contours
  (starting at, and spaced by, $2~\rm{}mJy$ per $11\arcsec$ beam after
  smoothing to the $20\arcsec$ beam marked by a circle). Markers
  indicate the position of L1148--IRS, its outflow, and of two of the
  positions searched for contraction motions. Note that L1148--IRS is
  offset from the nearby dust emission peak (by a marginally
  significant amount). Coordinates are given for
  J2000.\label{fig:layout-observations}}
\end{figure}

The IRAM 30m--telecope near Granada (Spain) was used to search for an
extended outflow from L1148--IRS in the ($J=2$--1) transition of
$\rm{}^{12}CO$. Data were taken on 2004 December 19 (single pixel
facility receiver; system temperature $\sim{}380~\rm{}K$ in
$T_{\rm{}mb}$--scale, mapped area $\sim{}2\arcmin{}\times{}2\arcmin{}$)
and 2005 July 19 to 21 (HERA multi--beam array; $\sim900~\rm{}K$,
$\sim{}4\arcmin{}\times{}4\arcmin{}$). The single--pixel maps were
already used in \citet{kauffmann2005:vellos}. The pointing corrections
in azimuth and elevation were below $3\arcsec$, and the focus
corrections below $0.45~\rm{}mm$, as checked by regular pointing and
focus checks at least once every 3 hours. The intensity scale was
converted to the main beam brightness temperature scale according to
the efficiencies reported on the IRAM
website\footnote{\path|http://www.iram.es/IRAMES/|}. Figure
\ref{fig:layout-observations} gives an overview of the layout of these
observations. The data were calibrated and analyzed using IRAM's CLASS
software package\footnote{\path|http://www.iram.fr/IRAMFR/GILDAS/|}.

%
%

%
%
%
%
%
%
%

In parallel to these outflow searches, we used a variety of other
tracers to probe the physical and chemical conditions in the dense
gas. This includes lines of $\rm{}N_2H^+$ (1--0) and $\rm{}N_2D^+$
(1--0), CS (2--1) and (3--2), $\rm{}C^{18}O$ (2--1), $\rm{}HCO^+$ and
$\rm{}H^{13}CO^+$ (1--0), and HCN (1--0). All observations were
obtained in frequency--switched mode using the single--pixel receivers
and the autocorrelator backends. The $\rm{}N_2H^+$ and $\rm{}C^{18}O$
lines were mapped on 2004 December 19 in the same
$\sim{}2\arcmin{}\times{}2\arcmin{}$ field for which $\rm{}^{12}CO$
data were obtained. Our $\rm{}N_2H^+$ and $\rm{}C^{18}O$ maps were
already used in \citet{kauffmann2005:vellos}. The remaining lines were
observed on 2005 July 19 to 21 in relatively deep integrations towards
the three positions near L1148--IRS sketched in Figure
\ref{fig:layout-observations} (chosen to probe L1148--IRS, its
immediate environment, and the neighboring dust core). Focus and
pointing uncertainties are as previously quoted for the outflow
search. More details are given in Table
\ref{app-tab-l1148:radio_obs}. CLASS was used to calibrate and exploit
the data. We removed low--order baselines and partially used Fourier
(FFT) filtering to further improve the baseline quality.

\subsection{KPNO 4m--telescope}
As part of a larger program to survey the c2d cores in the
near--infrared, deep $J$, $H$, and $K_{\rm{}s}$ observations of L1148
were obtained during September of 2004 using FLAMINGOS (Florida
Multi--object Imaging Near--IR Grism Observational Spectrometer;
\citealt{elston1998:flamingos}) on the 4m--telescope at Kitt Peak
National Observatory (KPNO). We use the data acquisition and reduction
strategies that have been described in \citet{huard2006:l1014}.

\subsection{Plateau de Bure
  Interferometer\label{sec:observations-pdbi}}
The Plateau de Bure Interferometer (PdBI) observed L1148--IRS on 6
occasions during the summer of 2006 (project P057; observed on May 13,
June 5 and 15, July 22, August 2, and September 1). Between 4 and 5
antennas of the array were used in the `D' configuration (i.e., the
most compact one). The $3~\rm{}mm$ and $1~\rm{}mm$ band receivers were
tuned to the ($J=1$--0) and (2--1) transitions of $\rm{}^{12}CO$,
respectively ($115.271202~\rm{}GHz$ and $230.538000~\rm{}GHz$;
\citealt{winnewisser1997:co}). All 8 units of the correlator took
data; for each receiver, 2 units were centered on the line to acquire
spectra at $0.1~\rm{}km\,s^{-1}$ and $0.2~\rm{}km\,s^{-1}$ resolution,
and 2 further units were symmetrically displaced from the line to
collect continuum data with a bandwidth of $320~\rm{}MHz$.  The
primary beam sizes were $20\arcsec$ and $40\arcsec$. The minimum
projected baseline in D configuration is $\sim{}20~\rm{}m$, and so
structures larger $\sim{}16\arcsec$, respectively $\sim{}33\arcsec$,
are not at all probed by the maps. L1148--IRS was in the phase center
of the array.

3C454.3 was typically used for bandpass calibration, while 1928+738
and 2037+511 served as phase calibrators. As usual for the PdBI,
MWC349 data yielded the initial flux calibration (assumed to have flux
densities of $1.12~\rm{}Jy$ and $1.7~\rm{}Jy$ in the $3~\rm{}mm$ and
$1~\rm{}mm$ bands, respectively). Based on this, the fluxes of the
other calibrators were fixed and iterated to obtain an optimum
calibration across all epochs. The flux calibration can be checked by
imaging the phase calibrator 2037+511 in those epochs where it is not
used to derive this calibration (i.e., is not held `fixed' in the
calibration routines). We find that the observations reproduce the
epoch--averaged mean flux density with maximum deviations of 7\% and
20\% in the $3 ~ \rm mm$ and $1 ~ \rm mm$ band, respectively.


The data were calibrated and imaged (including cleaning) using the
GILDAS software suite. Here, we use all epochs to construct maps
in the $3 ~ \rm mm$ band, 
but only the best 4 periods (i.e., excluding July 22 and August 2) are
taken in the $1 ~ \rm mm$ band. This yields beam sizes of
$4\farcs5 \times 3\farcs4$ (position angle east of north of $82\degr$)
and $2\farcs5 \times 1\farcs7$ (p.a.\ of $65\degr$) at $3 ~ \rm mm$ and
$1 ~ \rm mm$ wavelength, respectively. To improve the signal--to--noise
ratio, the spectra are sampled into bins of $0.3 ~ \rm km \, s^{-1}$
width. The continuum data are clipped in the spectral range to exclude
CO emission from our target source. This yields noise levels in the
$3 ~ \rm mm$ and $1 ~ \rm mm$ band spectral line data of
$23 ~ \rm mJy \, beam^{-1}$ and $46 ~ \rm mJy \, beam^{-1}$,
respectively, while we find $0.39 ~ \rm mJy \, beam^{-1}$ and
$1.0 ~ \rm mJy \, beam^{-1}$ for the continuum data.

\section{An Outflow from L1148--IRS}
Previous analysis failed to establish that L1148--IRS (\textit{i}) is a
YSO and (\textit{ii}) is physically associated with the L1148
region. Here, we show that the source drives a weak small--scale CO
outflow ($\lesssim 5\arcsec$, about $1800~\rm{}AU$ in
projection). This confirms the hypotheses (\textit{i}) and
(\textit{ii}).\label{sec:outflow}

\subsection{A Compact CO Outflow\label{sec:outflow-compact_CO}}
\subsubsection{Single--dish Observations}
We initially used the IRAM 30m--telescope to search for an outflow in
the $\rm{}^{12}CO$ (2--1) line. This search failed
\citep{kauffmann2005:vellos}. It is briefly summarized below.

\begin{figure}
\includegraphics[height=0.8\linewidth,angle=-90]{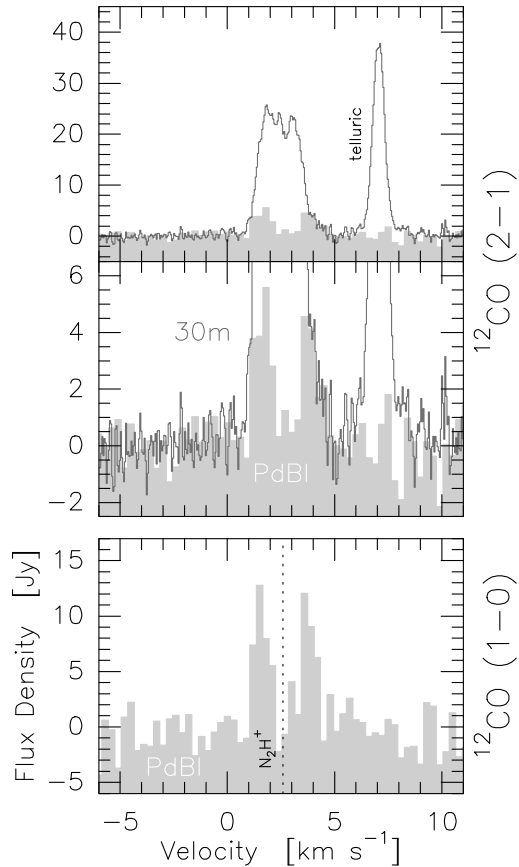}
\caption{CO spectra towards L1148--IRS. The panels show $\rm ^{12}CO$
  ($J=2$--1) data (top) and ($J=1$--0) spectra (bottom). For the
  (2--1) transition, the solid line presents the frequency--switched
  IRAM 30m--telescope spectrum ($11\arcsec$ beam) taken towards
  L1148--IRS. The signal near $7 ~ \rm km \, s^{-1}$ velocity is of
  telluric origin and thus not relevant here. The shaded histograms
  portray the data acquired with the Plateau de Bure
  Interferometer. These spectra are averaged over source--centered
  apertures of $3\arcsec$ and $10\arcsec$ radius for the (2--1) and
  (1--0) lines, respectively. The systemic velocity derived from $\rm
  N_2H^+$ observations (Section \ref{sec:previous-work}, Appendix
  \ref{app-sec:velocity-field}) is marked by a dashed
  line.\label{fig:outflow-spectrum}}
\end{figure}

The single--pixel facility receiver was used to map an area of
$2\arcmin{}\times{}2\arcmin$ size with $30 \arcsec$ spacing in
frequency--switched mode (layout indicated in Figure
\ref{fig:layout-observations}). The top panels of Figure
\ref{fig:outflow-spectrum} show an example spectrum from this search:
emission from the dense core (as well as ``telluric''
emission\footnote{The LSR velocity of the telescope can, e.g., be
  found from the calculator provided on the homepage of the \emph{Far
    Ultraviolet Spectroscopic Explorer} (switched to topocentric
  coordinates),
  \path|http://fuse.pha.jhu.edu/support/tools/vlsr.html|, and using
  the ``source'' command within the ASTRO program of the GILDAS
  software package (\path|http://www.iram.fr/IRAMFR/GILDAS/|). The LSR
  velocity of telluric lines is the negative of the telescope's LSR
  velocity.} from Earth's atmosphere) are detected, but no obvious
outflow signal was found (neither in the form of line wings, nor in
terms of clear spatial trends in emission). However, Fourier
components, as well as baselines of $2^{\rm nd}$ order, had to be
removed during data reduction. Thus, these observations were not
sensitive to outflows manifesting in wide and faint emission lines.

Therefore, we later raster--mapped a field of about
$4\arcmin{}\times{}4\arcmin$ size with $24\arcsec$ spacing using the
HERA multi--beam array in position--switched mode (indicated in Figure
\ref{fig:layout-observations}). These data have very flat
baselines. Unfortunately, the reference position was not free of
emission. Therefore, the data are severely contaminated within about
$\pm{}2~\rm{}km{}\,{}s^{-1}$ from systemic velocity. After smoothing
to $0.5~\rm{}km{}\,{}s^{-1}$ velocity resolution, these spectra have a
noise level of $0.8~\rm{}Jy{}\,beam^{-1}$ in the main beam brightness
scale. Still, no outflow emission was found. Also the mean
  spectrum averaged over all map pixels does not reveal any CO emission beyond
$\pm{}2~\rm{}km\,s^{-1}$ velocity offset (at a noise level of
$54~\rm{}mJy{}\,beam^{-1}$).

Figure \ref{fig:outflow-spectrum} compares the single--dish
observations with the interferometer--detected outflow described
below. For this comparison, the PdBI spectra in the $\rm{}^{12}CO$
(2--1) and (1--0) transitions are integrated over source--centered
apertures of $3\arcsec$ and $10\arcsec$ radius, respectively. The data
show that the single--dish observations are sensitive enough to detect
the outflow. The outflow emission is, however, massively confused with
the dense core emission. This is similar to the situation in L1014
\citep{bourke2005:l1014-outflow}.


\subsubsection{Interferometric Observations}
\begin{figure}
\includegraphics[height=\linewidth,angle=-90]{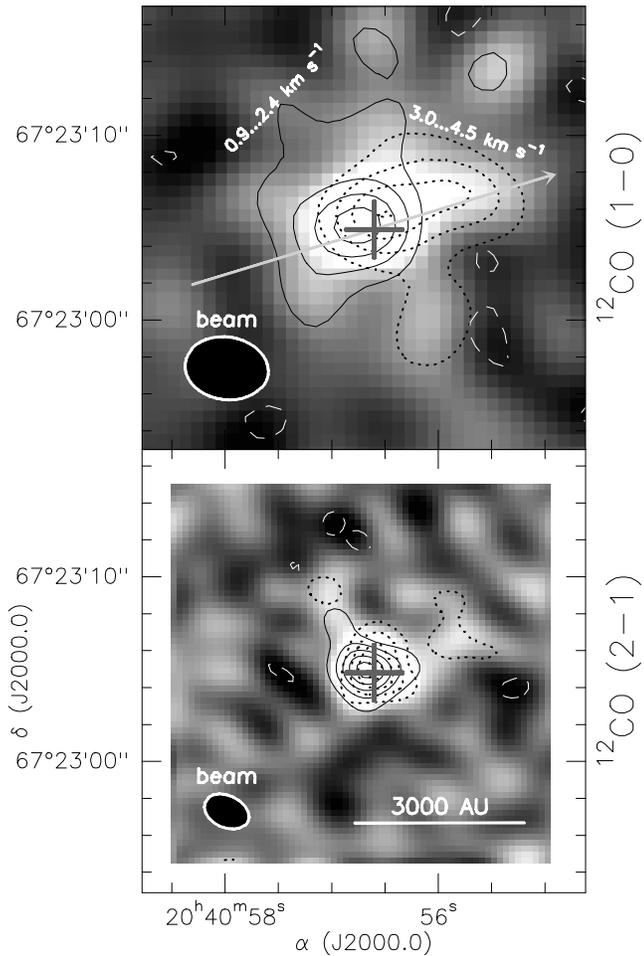}
\caption{The outflow lobes for both $\rm ^{12}CO$
  transitions. The greyscale map presents the emission
  integrated over the $0.9 ~ {\rm to} ~ 4.5 ~ \rm km \, s^{-1}$
  velocity interval. Solid contours outline the emission of the
  blue-shifted lobe at $0.9 ~ {\rm to} ~ 2.4 ~ \rm km \, s^{-1}$
  velocity, while dotted contours delineate the red-shifted
  lobe at $3.0 ~ {\rm to} ~ 4.5 ~ \rm km \, s^{-1}$. These contours
  start at, and are spaced by, the noise level times a factor
  $\pm 3$; they are dashed for areas with negative
  emission. The central cross marks the position of
  L1148--IRS. Ellipses give the beam sizes at half
  sensitivity. The arrow marks the location and orientation of
  the position--velocity cut presented in Figure
  \ref{fig:outflow-pv}.\label{fig:lobes}}
\end{figure}

\begin{figure}
\includegraphics[height=\linewidth,bb=272 67 470 420,clip,angle=-90]{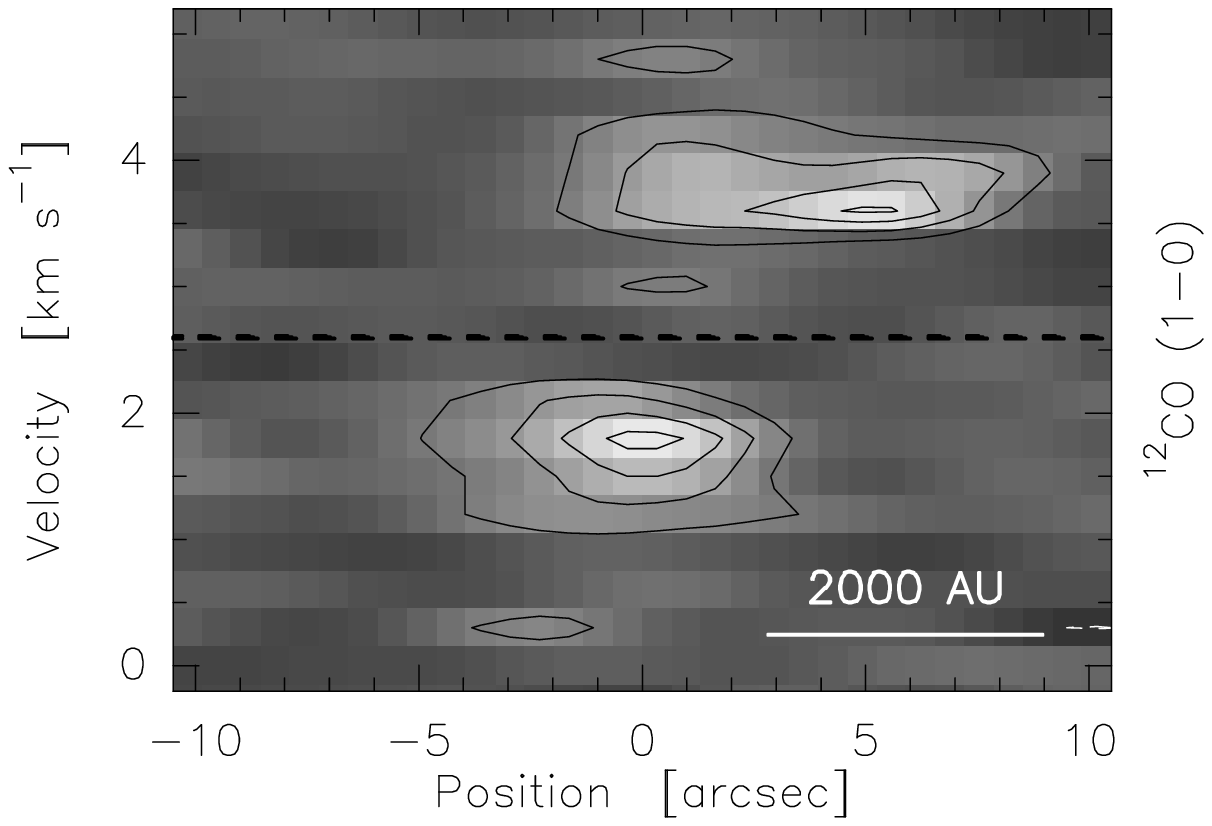}
\includegraphics[height=\linewidth,bb=272 67 506 420,clip,angle=-90]{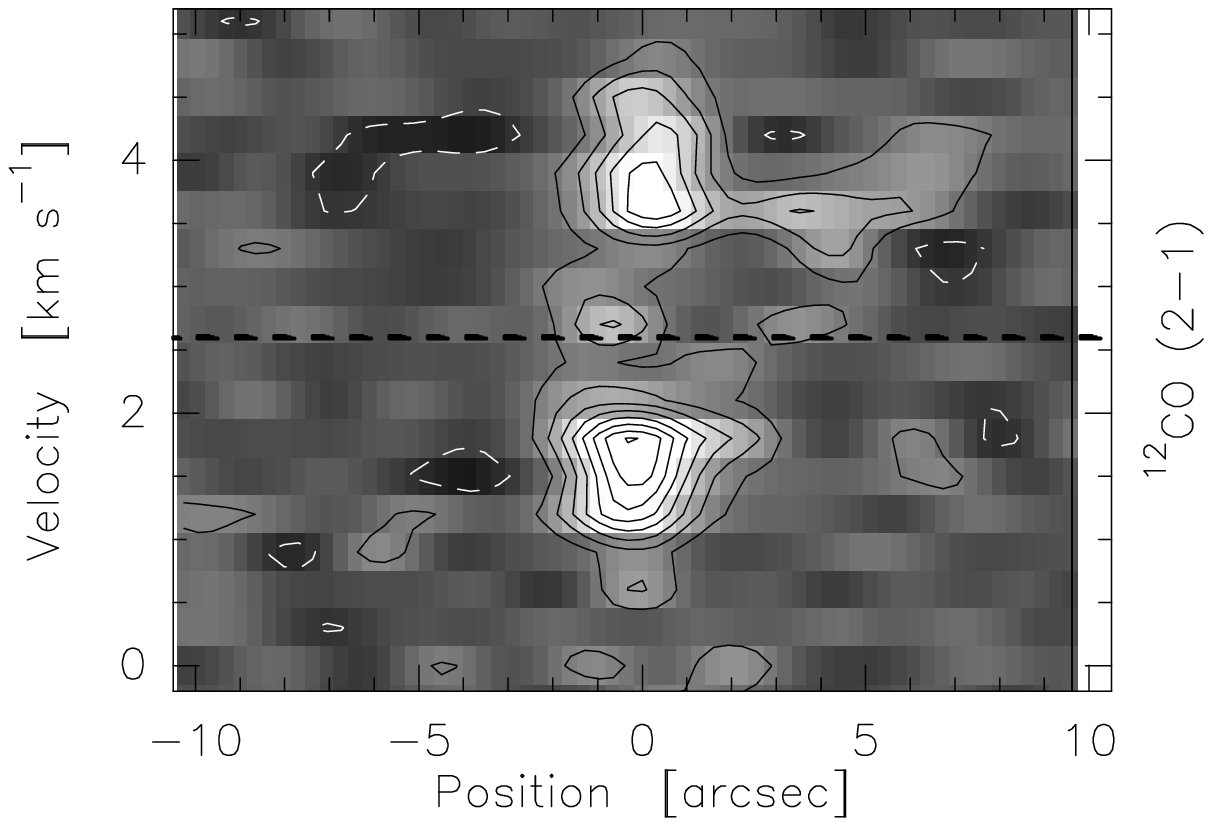}
\caption{Position--velocity diagrams along the trajectory indicated in
  Figure \ref{fig:lobes}. The spatial origin coincides with
    L1148--IRS, and the offset increases in the direction of the arrow
    drawn in Figure \ref{fig:lobes}. Contours start at, and are spaced by,
  the noise level times a factor $\pm 2$. Dashed lines are drawn
  for negative signal, and solid contours are used
  otherwise. The horizontal dashed line marks the systemic
  velocity derived from $\rm N_2H^+$ observations (Section
  \ref{sec:previous-work}, Appendix \ref{app-sec:velocity-field}).\label{fig:outflow-pv}}
\end{figure}

The non--detection with single--dish telescopes did not rule out the
existence of a compact outflow from L1148--IRS, such as the one seen in
L1014 \citep{bourke2005:l1014-outflow}. We therefore also used IRAM's
Plateau de Bure Interferometer (PdBI) to observe the (1--0) and (2--1)
transitions of $\rm{}^{12}CO$. These observations reveal an outflow
from L1148--IRS. As we detail below, it manifests in two symmetric
velocity components (Figures \ref{fig:outflow-spectrum} and
\ref{fig:outflow-pv}) that form a bipolar outflow system (Figure
\ref{fig:lobes}).

In the PdBI--derived data cubes, significant $\rm{}^{12}CO$ emission is
detected in the $1.2~{\rm{}to}~4.2~\rm{}km\,s^{-1}$ velocity
range. (``Significant'' means that, in a given
position--position--velocity cell, the emission of both $\rm{}^{12}CO$
transitions exceeds the noise level by a factor $\ge{}4$. The velocity
channels are spaced by $0.3~\rm{}km\,s^{-1}$.) 
The $\rm{}N_2H^+$--derived systemic velocity is
  $2.60~\rm{}km\,s^{-1}$ (Section \ref{sec:previous-work}, Appendix
  \ref{app-sec:velocity-field}). Relative to
  this we obtain maximum projected velocity offsets of
$-1.4~{\rm{}to}~+1.6~\rm{}km\,s^{-1}$ for the $\rm{}^{12}CO$ outflow
emission. We approximate this as $\pm{}1.5~\rm{}km\,s^{-1}$ in the
following. Correspondingly, we use velocity ranges of
$0.9~{\rm{}to}~2.4~\rm{}km\,s^{-1}$ (blue--shifted lobe) and
$3.0~{\rm{}to}~4.5~\rm{}km\,s^{-1}$ (red--shifted lobe; we exclude the
central channel at $2.7~\rm{}km\,s^{-1}$ for clarity) to calculate the
velocity--integrated emission of the lobes.

Figure \ref{fig:lobes} presents integrated intensities in order to
explore the structure of the lobes. These are more apparent in the
$\rm{}^{12}CO$ (1--0) data. In this line, the blue--shifted lobe is
shorter than the red--shifted one. The extent of the latter lobe is
$5\farcs5\pm{}0\farcs3$, or $(1800\pm{}100)~\rm{}AU$ in projection; to
measure this, we find the most elongated contour (at a signal--to--noise
level of $9.5\pm{}1.0$), drop to 50\% of this intensity, measure this
contour's maximum offset from L1148--IRS, and subtract the beam radius
linearly. Manual fitting of the bright and elongated outflow emission
with a straight line (Figure \ref{fig:lobes}) gives a most likely
position angle of $107\degr{}\pm{}22\degr$. (To calculate the
uncertainty, we derive the position angles when displacing the fitted
line by half a beam at $5\farcs5$ offset from L1148--IRS.) This fit is
guided by the fact that L1148--IRS resides right between the respective
intensity peaks of the two lobes; this suggests an outflow direction
similar to the position angle of the line connecting the
peaks. Unfortunately, though, the emission farther away from the star
($\gtrsim{}5\arcsec$) shows no clear structure. This leaves the option
of a wide--angle outflow propagating in a direction significantly
different from the one obtained above.

Figure \ref{fig:outflow-spectrum} shows spectra of the outflow
emission.  Gaussian
fits to the spectra, and comparison to the $\rm N_2H^+$ systemic
velocity, yield projected velocity offsets of $-1.0$ and
$+1.2~\rm{}km\,{}s^{-1}$ for the $\rm{}^{12}CO$ (1--0) line,
respectively $-0.9$ and $+1.2~\rm{}km\,{}s^{-1}$ for the (2--1)
line. We approximate this as $\pm{}1.0~\rm{}km\,{}s^{-1}$ in the
following. Gaussian line fitting of the PdBI spectra gives
velocity--integrated flux densities of $21.0~\rm{}Jy\,{}km\,{}s^{-1}$
and $8.9~\rm{}Jy\,{}km\,{}s^{-1}$ (15\% of the single-dish flux
density) for the (1--0) and (2--1) lines, respectively.

The position--velocity diagrams presented in Figure
\ref{fig:outflow-pv} show that the lobes are kinematically
well--separated along the entire outflow axis. Some faint outflow
features can possibly be traced to velocity offsets as large as
$\pm{}2.5~\rm{}km\,{}s^{-1}$. The $\rm{}^{12}CO$ (2--1) panel reveals
that the knot at $+5{}\arcsec$ offset is detected in several velocity
channels, and is therefore probably real. The elongated lobe seen in
the (1--0) maps appears to form a counterpart of this knot.

The CO emission cannot come from a circumstellar disk. Keplerian
motion at $1 ~ \rm km \, s^{-1}$ speed and $1800 ~ \rm AU$ radius
would require a central mass of $2 \, M_{\sun}$. This significantly
exceeds the estimated mass of the central star (Section
\ref{sec:stage-mass}) and of the  dense
  core material immediately surrounding the YSO (Section
\ref{sec:previous-work}). The emission must thus come from an outflow.

There is a remote chance that the compact CO emission is not from an
outflow, but is just an artifact of the interferometer--induced
spatial filtering of the dense core emission. We cannot rule this
option out. However, as we will show in Section
\ref{sec:YSO-confirmation}, the close association of CO emission and
L1148--IRS alone is sufficient to reliably demonstrate the
protostellar nature of L1148--IRS.

\subsection{Outflow Mass and Kinematics}
To derive the outflow mass, we assume the $\rm^{12}CO$ emission to be
optically thin, adopt an $\rm{}H_2$--to--$\rm{}^{12}CO$ abundance ratio
of $10^4$, a mean mass per $\rm{}H_2$ molecule of 2.4 H--masses, and
use Eq.\ (A4) of \citet{bourke1997:outflow-bhr71} to calculate the
column densities (for $\rm^{12}CO$, we use a permanent dipole moment
of $0.11~\rm{}D$ and a rotational constant of $57.636~\rm{}GHz$). We
adopt an excitation temperature of $20~\rm{}K$. This is a bit
  more than canonical dense core temperatures $\sim{}10~\rm{}K$, but
  less than the few $10^2~\rm{}K$ observed in outflows of much higher
  velocity (see, e.g., \citealt{santiago-garcia2009:iras04166} for a
  summary). This temperature uncertainty introduces a moderate mass
  uncertainty: temperatures in the range $10~{\rm{}to}~50~\rm{}K$
would give masses different by less than a factor 2. We obtain outflow
mass estimates, $m_{\rm{}out}$, of $1.3\times{}10^{-3}\,M_{\sun}$ and
$3.9{}\times{}10^{-4}\,M_{\sun}$ for the $\rm{}^{12}CO$ (1--0) and
(2--1) observations, respectively. Correction for typical
$\rm{}^{12}CO$ outflow optical depths of 2 to 5
\citep{levreault1988:outlow-search} would increase the masses by
similar factors. For the (2--1) line, comparison with single--dish data
implies a total flux loss of a factor 7 (Figure
\ref{fig:outflow-spectrum}). This could lead to underestimation
  of the outflow mass by similar factors.

\begin{table}
  \caption{Outflow properties. The quoted uncertainties refer to the
    range in outflow masses derived from the different $\rm{}^{12}CO$
    transitions.\label{tab:outflow-properties}}
\begin{tabular}{llllllllll}
\hline
Property & Definition & Value\\ \hline
size (radius) \vspace{2ex} & $\displaystyle \frac{r}{\rm{}AU}$ & $1800$ \\
velocity \vspace{2ex} & $\displaystyle \frac{v_{\rm{}out}}{\rm{}km\,{}s^{-1}}$ &
 $\left|{}\pm{}1\right|$\\
mass \vspace{1ex} & $\displaystyle \frac{m_{\rm{}out}}{10^{-3}\,M_{\sun}}$ &
$0.4~{\rm{}to}~1.3$\\
momentum \vspace{1ex} & $\displaystyle
\frac{m_{\rm{}out}\cdot{}v_{\rm{}out}}{10^{-3}\,{}M_{\sun}\,{}\rm{}km\,{}s^{-1}}$
& $0.4~{\rm{}to}~1.3$\\
energy \vspace{1ex} & $\displaystyle \frac{m_{\rm{}out}\cdot{}v_{\rm{}out}^2/2}{10^{-4}\,{}M_{\sun}\,{}\rm{}km^2\,{}s^{-2}}$ &
 $0.2~{\rm{}to}~0.6$ \\
mechanical luminosity \vspace{1ex} & $\displaystyle \frac{m_{\rm{}out}\cdot{}v_{\rm{}out}^3/[2\,{}r]}{10^{-5}\,{}L_{\sun}}$ &
 $0.4~{\rm{}to}~1.3$ \\
force \vspace{1ex} & $\displaystyle \frac{m_{\rm{}out}\cdot{}v_{\rm{}out}^2/r}{10^{-7}\,{}M_{\sun}\,{}\rm{}km\,{}s^{-1}\,{}yr^{-1}}$ &
  $0.5~{\rm{}to}~1.5$ \\
dynamical age \vspace{1ex} & $\displaystyle \frac{r/v_{\rm{}out}}{10^3~\rm{}yr}$ & $8.5$\\
rate & $\displaystyle \frac{m_{\rm{}out}\cdot{}v_{\rm out}/r}{10^{-7}\,{}M_{\sun}\,{}\rm{}yr^{-1}}$ & $0.5~{\rm{}to}~1.5$\\ \hline
\end{tabular}
\end{table}

We adopt a characteristic outflow velocity of
$v_{\rm{}out}={}\left|{}\pm{}1~\rm{}km\,{}s^{-1}{}\right|$, and a
radial size of $r=1800\,{}\rm{}AU$. These are, of course, projected
properties, which in principle must be corrected for the inclination
angle. This angle is, unfortunately, not known, which introduces
systematic uncertainties into our analysis. Using the formalism of
\citet{cabrit1990:outflow-co} and \citet{bourke1997:outflow-bhr71}, we
derive the properties listed in Table
  \ref{tab:outflow-properties}.  The quoted uncertainties refer to
the range in outflow masses derived from the different $\rm{}^{12}CO$
transitions. Given their linear dependence on $m_{\rm{}out}$, the
outflow properties also suffer from the systematic uncertainties
affecting mass estimates.\medskip

\noindent{}The properties of L1148--IRS and its outflow are unusual,
but not extreme for an embedded YSO. L1148--IRS is less luminous than
any source in the comprehensive \citet{wu2004:outflows} outflow
compilation. It also has an outflow mass, mechanical luminosity, and
force smaller than observed for all (but 4, allowing for
uncertainties) of the 292 outflow systems with known luminosity in the
\citeauthor{wu2004:outflows} study. These trends do not merely reflect
the low luminosity of L1148--IRS; for L1148--IRS, the above properties
are still 1 to 2 magnitudes below what is derived when extrapolating
typical outflow properties from \citet{wu2004:outflows} to the
internal luminosity of L1148--IRS. These deviations are, however,
marginally within the scatter observed for other YSOs, and the
L1148--IRS outflow is thus not extreme. The outflow survey of embedded
YSOs by \citet{bontemps1996:outflows} can be used for another
comparison. Based on this study, the force of the L1148--IRS outflow is
consistent with what is expected for class I sources of comparable
internal luminosity. The YSO envelope mass of L1148--IRS does, however,
exceed the typical one expected for the observed outflow force by a
factor $\sim 10$.

\citet[][ IRAM04191]{andre1999:iram04191},
\citet[][ L1014--IRS]{bourke2005:l1014-outflow} and
\citet[][ ISO--Oph 102]{phan-bao2008:bd-outflow} report outflow
properties for other VeLLOs and brown dwarfs. The outflow of L1148--IRS
is similar to those of L1014--IRS and ISO--Oph 102 in terms of
velocity (all have $\sim 1 ~ \rm km \, s^{-1}$), but its mass
($\sim 10^{-3} \, M_{\sun}$ vs.\
$\sim 10^{-5} \, M_{\sun}$, without mass corrections) and size
($1800 ~ \rm AU$ vs.\ $\le 1000 ~ \rm AU$) appear to be larger. This
is small, though, compared to the $\sim 10 ~ \rm km \, s^{-1}$
velocity range and $\sim 15000 ~ \rm AU$ extent of IRAM04191.

\subsection{Absence of Compact Dust Emission\label{sec:dust-disk}}
Compact dust emission, such as emitted by circumstellar disks, is not
detected towards L1148--IRS. At $1~\rm{}mm$ and $3~\rm{}mm$ wavelength,
we observe intensities of $0.23~\rm{}mJy\,beam^{-1}$ and
$0.56~\rm{}mJy\,beam^{-1}$, respectively. This is insignificant
compared to the noise levels reported in Section
\ref{sec:observations-pdbi}. If we add the noise level times a factor
3, we obtain upper flux density limits of $<3.2~\rm{}mJy$ and
$<1.7~\rm{}mJy$, where we assume that the dust emission should be more
compact than the beam.

These upper limits are in line with, e.g., interferometric continuum
observations of L1014--IRS \citep{bourke2005:l1014-outflow}. To derive
mass limits from these flux densities, we adopt the framework laid out
by \citet{kauffmann2008:mambo-spitzer}. Assuming dust with
temperatures $\ge{}10~\rm{}K$, this yields mass limits
$<0.02\,M_{\sun}$ and $<0.13\,M_{\sun}$, respectively.

%

\subsection{Near--Infrared Nebulosity}
\begin{figure}
\includegraphics[height=\linewidth,angle=-90]{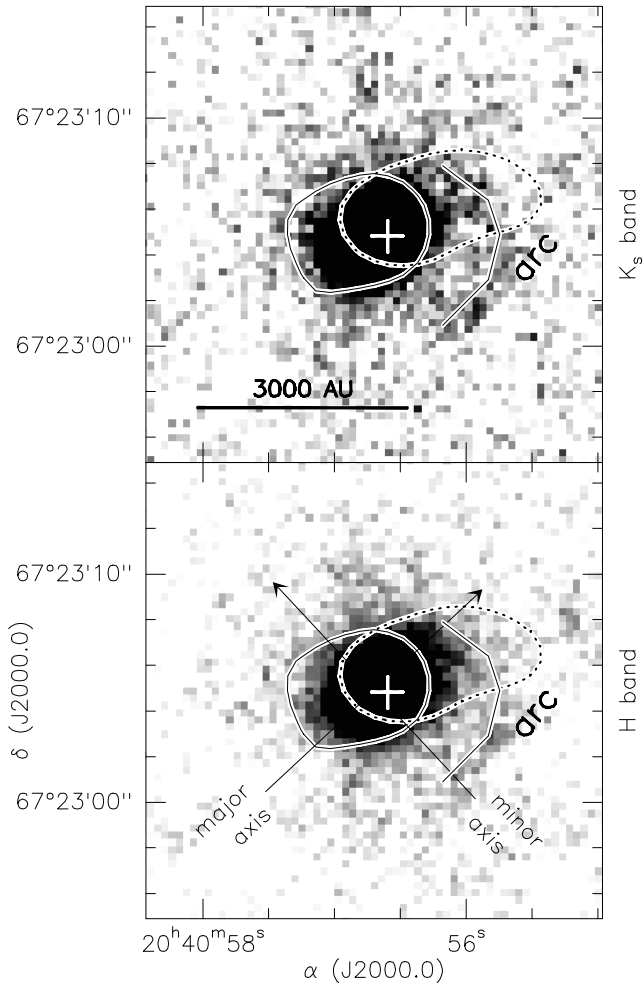}
\caption{Images of L1148--IRS in the $H$ (bottom panel) and $K_{\rm s}$
  bands (top panel) as seen by FLAMINGOS, shown as their
  negative. Contours delineate the $\rm ^{12}CO$ (1--0) outflow lobes
  for signals exceeding the noise by a factor 6 (see Figure
  \ref{fig:lobes}). The arrows indicate positions and directions of
  the cuts through the intensity distribution that are discussed in
  the text. The images reveal that L1148--IRS is elongated, with the
  major axis at position angle $\approx 133 \degr$. In addition there
  is an arc--like extended feature about $5 \arcsec$ to the west of
  L1148--IRS. To facilitate comparison between both images, the arc is
  indicated by a solid tracer line.\label{fig:L1148_HK}}
\end{figure}

The near--infrared (NIR) FLAMINGOS imaging observations presented in
Figure \ref{fig:L1148_HK} reveal elongated nebulosity towards
L1148--IRS. This nebulosity is likely related to outflows from the
star.

L1148-IRS is detected in all bands, but we only use the
two bands with the best data (i.e., $H$ and $K_{\rm{}s}$) for our
analysis. Fits of elongated Gaussians to the $H$--band image (which
has the higher SNR) give a position angle (east of north) of
$133\degr$ for the major axis. Along this axis, $H$--band emission
from L1148--IRS above 10\% of the peak intensity has a width of
$4\farcs{}1$ ($3\farcs{}7$ in the $K_{\rm{}s}$ band), while nearby
stars in the image have a width of only $2\farcs{}5$ ($2\farcs{}7$ for
the $K_{\rm{}s}$ band) for this position angle. Repetition of this
procedure for the minor axis, and removal of the instrument--induced
source width at 10\% peak intensity, suggests an intrinsic extension
of order $3\farcs{}2\times{}2\farcs{}4$ or
$1040\times{}780~\rm{}AU^2$. Both bands reveal an arc--like feature
about $5\arcsec$ to the west of L1148--IRS. This could be related to
the aforementioned western blob in the red--shifted $\rm ^{12}CO$
(2--1) outflow lobe.

The CO--detected lobes and the NIR nebulosity are well aligned
($133\degr$ vs.\ $107\degr \pm 22\degr$).  The nebulosity thus seems
to trace structures such as an outflow--blown cavity.

\subsection{L1148--IRS as a Confirmed
  VeLLO\label{sec:YSO-confirmation}}
As pointed out by \citet{dunham2008:vellos}, confirmation of a VeLLO
candidate requires (\textit{i}) that the internal luminosity is below
$0.1 \, L_{\sun}$, and (\textit{ii}) proof of embeddedness in a dense
core. In the characterization framework of \citet{dunham2008:vellos},
the ouflow discovery furnishes the latter. Following the terminology
of \citeauthor{dunham2008:vellos}, L1148--IRS is thus a ``confirmed
embedded low--luminosity protostar'' (their `group 1').

However, given the relatively weak emission, low velocity, and small
lobe separation in L1148--IRS, one may doubt that the compact CO
emission is really coming from an outflow. As we show here, though,
the proximity of compact CO emission and IR source independently
demonstrates that L1148--IRS is physically associated to the dense
core. Then, L1148--IRS must still be deeply embedded, since there is no
other way to produce the Spitzer MIPS emission at the observed
intensities (Section \ref{sec:sed}).

%
%
%
%

We establish the association by computing the chance alignment
probability between background sources (i.e., galaxies) and the
$\rm{}^{12}CO$ emission. A background source density
$\sim{}2000~\rm{}sr^{-1}$ is to be expected for $24~\rm{}\mu{}m$
sources brighter $70~\rm{}mJy$
\citep{papovich2004:galaxy_counts}. L1148--IRS and the $\rm{}^{12}CO$
(2--1) emission peaks are aligned to within $1\arcsec$ distance
(Figure \ref{fig:lobes}). For an individual cloud, the probability for
chance alignment is thus $\pi{}\,(1\arcsec)^2\cdot
2000~{\rm{}sr^{-1}}=1.5\times{}10^{-8}$. For the entire c2d MAMBO
survey, with $\sim{}110$ dust emission peaks, the probability of one
chance alignment is $1.6\times{}10^{-6}$ (assuming that every dust
emission peak contains interferometer--detected CO structures like
those presented here). There is no realistic option that dense core
and L1148--IRS are not physically associated.

\section{The Spectral Energy Distribution of L1148--IRS}
The outflow observations establish the protostellar nature of
L1148--IRS. However, we still need to characterize the embedded
infrared source. Here, we gather photometry of this object, sketch a
model for the analysis of the observations, and discuss the properties
derived using this model.\label{sec:sed}

\subsection{Observations\label{sec:sed-observations}}
\begin{figure}
\includegraphics[width=\linewidth,angle=0]{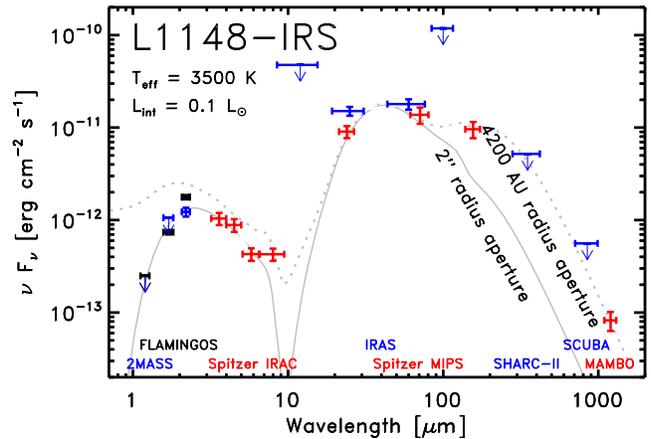}
\caption{The spectral energy distribution of L1148--IRS. Bars represent
  flux density measurements with their associated uncertainty and
  filter width, while arrows indicate upper limits. Solid and
  dashed lines describe a dense core model
  containing a point source with $0.1 \, L_{\sun}$ internal luminosity
  and $3500 ~ \rm K$ effective temperature; the solid line holds for
  an aperture of $2\arcsec$ radius, which is used to model the
  observations at $\le 10 ~ \micron$ wavelength, while the dashed line
  refers to an aperture of $4200 ~ \rm AU$ radius, as applied to the
  observations at larger wavelengths. This is detailed in Section
  \ref{sec:SED_modelling}.\label{fig-l1148:L1148_SED}}
\end{figure}

Figure \ref{fig-l1148:L1148_SED} presents the spectral energy
distribution (SED) of L1148--IRS. These data are assembled from the
2MASS point source catalogue \citep{skrutskie2006:2mass}, our
FLAMINGOS observations, the IRAS faint source catalogue
\citep{moshir1990:iras_faint_source_cat}, the final release of the c2d
source catalogue \citep{evans2007:c2d-delivery}, our cores2deeper
Spitzer project (PID 20386), our deep MIPS $160 ~ \rm \mu m$ survey of
cores (PID 30384), SHARC--II imaging by \citet{wu2007:c2d_sharc-ii},
the SCUBA legacy source catalogue
\citep{difrancesco2008:scuba_catalogue}, and our c2d MAMBO survey
\citep{kauffmann2008:mambo-spitzer}. Appendix \ref{app:photometry}
discusses how we consolidate these data from different origins. Table
\ref{tab-l1148:L1148_SED} lists the flux densities in detail.

Integration of the SED yields bolometric luminosities and temperatures
of $0.12\,L_{\sun}$ and $110~\rm{}K$, respectively. We derive these
parameters using power--laws to interpolate between the detected flux
densities. We further assume flux densities scaling with wavelength as
$\lambda^4$ for $\lambda>350~\rm{}\mu{}m$
\citep{kauffmann2008:mambo-spitzer}. Propagation of the flux--weighted
observational uncertainties gives uncertainties $<20\%$ in luminosity,
and $\pm{}15~\rm{}K$ in temperature. This is in agreement with the
L1148--IRS characterization by
\citet{kauffmann2005:vellos,kauffmann2008:mambo-spitzer}, but a bit
warmer and fainter than what is found by
\citeauthor{kirk2009:cepheus-flare}
(\citeyear{kirk2009:cepheus-flare}; probably because they use MAMBO
fluxes for larger apertures).

\subsection{Modeling\label{sec:SED_modelling}}
We model the SED of L1148--IRS as a spherical dense core that contains
a point source and is externally illuminated by the interstellar
radiation field (ISRF). The point source is assumed to emit like a
black body of luminosity $L_{\rm{}int}$ and temperature
$T_{\rm{}eff}$; we refer to these as the internal luminosity and the
stellar effective temperature. The ISRF is taken from
\citet{perault1987:thesis}. We reduce its intensity at the core
surface by a factor 4.  This reflects shielding by extended material
of a total (surface to surface) extinction of $3~\rm{}mag$, and is
necessary to simultaneously match the MAMBO and MIPS3 observations. We
adopt dust opacities as \citet{weingartner2001:dust-size} predict for
their ``case B'' and a total--to--selective extinction ratio of
$R_V=5.5$ (since this model provides a good fit to observed dust
opacities at wavelengths $\lesssim{}10~\rm{}\mu{}m$; e.g.,
\citealt{indebetouw2005:extinction}). Model flux densities are
calculated using a code kindly provided by E.\ Kr\"ugel (e.g.,
Appendix A of \citealt{pillai2006:g11}).

The dense core's density profile, $\varrho(r)$, is modeled as a
continuous step--wise power law. We truncate the sphere at an outer
radius of $32500~{\rm{}AU}\approx0.16~\rm{}pc$, corresponding to
$\approx{}100\arcsec$. The density is assumed to be negligible within
the inner truncation radius, $r_{\rm{}in}$. For most radii, the slope
of the density profile is set to
${\rm{}d}\,\ln(\varrho)/{\rm{}d}\,\ln(r)=-2$. However, good matches of
the SED require a flatter density law in some inner part of the
sphere. We do this by introducing a constant density between the inner
and outer transition radii, $r_{\rm{}trans}^{\rm{}in}$ and
$r_{\rm{}trans}^{\rm{}out}$. Our fits use
$r_{\rm{}trans}^{\rm{}out}=2\times{}10^{14}~{\rm{}m}\approx{}1340~\rm{}AU$
(i.e., $8\farcs25$ diameter), since this is large enough to provide
good SED fits (as demonstrated by the results below) but is compact
enough to be consistent with the steep dust emission intensity profile
that rules out extended regions of constant density (Section
\ref{sec:previous-work}). With these assumptions, the density
profile is fully defined once the density at the outer truncation
radius, $\varrho_{\rm{}out}$, is specified. We fix the latter by
requiring the model to exactly match the MAMBO flux density
measurement. This eventually leaves two envelope parameters (i.e.,
$r_{\rm{}in}$ and $r_{\rm{}trans}^{\rm{}in}$) and two source
parameters (i.e., $L_{\rm{}int}$ and $T_{\rm{}eff}$)
unconstrained.\medskip

\noindent{}At wavelengths $<10~\rm{}\mu{}m$, we attempt to minimize
the mean squared relative difference between the observed and
predicted flux densities (i.e., the difference between model and
observation, divided by the observation; this property,
$\chi_{\rm{}rel}^2$, takes the role of the classical $\chi^2$).  We do
not normalize the differences by the observational flux density
uncertainties (as usually done to derive $\chi^2$) because the
predicted flux densities do sensitively depend on the adopted geometry
(e.g., \citealt{whitney2003:evol_sequence}). In this situation, the
deviations between good models and observed flux densities are not
dominated by observational uncertainties. We further require that a
valid model does strictly not exceed the $J$--band limits. The
synthetic observations are calculated for an aperture of $2\arcsec$
radius, and they are corrected for an intensity background determined
between $5\arcsec$ and $10\arcsec$ radius. This is used to capture the
effect of the background subtraction in the 2MASS, FLAMINGOS, and IRAC
observations.

At wavelengths $> 10 ~ \rm \mu m$, we require the model to strictly
match the observations within their observational uncertainties
(except for all IRAS and the MIPS3 bands, which are too
uncertain). This appears appropriate since in this wavelength domain,
deviations between good model and observation should be dominated by
observational uncertainties. We use an aperture of $4200 ~ \rm AU$
radius to calculate synthetic observations. This is exactly the
aperture used in MAMBO observations (Section
\ref{sec:previous-work}). The exact aperture size does not matter
for the MIPS1 and MIPS2 bands, since the emission originates (in the
model) at radii smaller $4200~\rm{}AU$.

The PdBI--derived upper limits for dust continuum emission are not
useful for our analysis of the SED of the L1148 core on large spatial
scales. Specifically, the interferometer probes radii $\le{}4\farcs5$,
i.e.\ $\le{}730~\rm{}AU$, while the analysis above considers scales as
large as $4200~\rm{}AU$. The PdBI maps are only useful to examine the
(absence of) disks around L1148--IR, as done in Sec.\
\ref{sec:dust-disk}.

\subsection{Results\label{sec:sed--results}}
\begin{figure}
  \includegraphics[height=\linewidth,angle=-90]{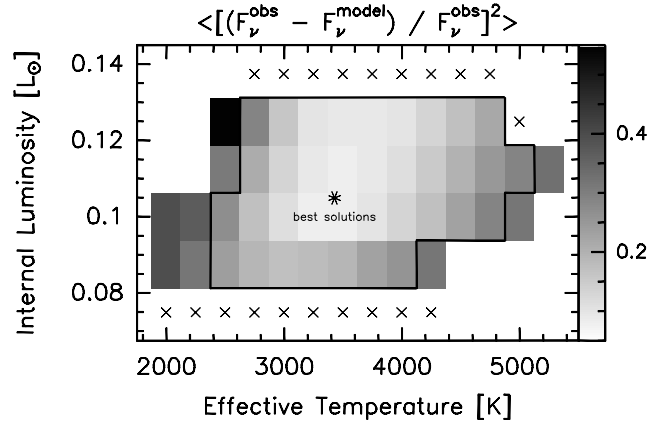}
  \caption{Quality of SED fits in dependence on the model effective
    temperature and internal luminosity. Only cells with shading or
    crosses have been examined. Grey shading gives
    $\chi_{\rm{}rel}^2$, the mean squared relative deviation between
    SED model and observations at wavelengths
    $<10~\rm{}\mu{}m$. Crosses are drawn if no model can fit all of
    the SED constraints. As explained in the main text, only solutions
    with $\chi_{\rm{}rel}^2<0.3$ are deemed acceptable. These
    solutions are enclosed by a black line. The modeling thus
    constrains the internal luminosity to
    $0.08\lesssim{}L_{\rm{}int}/L_{\sun}\lesssim 0.13$ and the
    effective temperature to
    $2500\lesssim{}T_{\rm{}eff}/{\rm{}K}\lesssim{}5000$.\label{fig-l1148:IR-properties}}
\end{figure}

Figure \ref{fig-l1148:IR-properties} presents the smallest mean
squared relative deviation found (by varying $r_{\rm{}in}$ and
$r_{\rm{}trans}^{\rm{}in}$) for a given pair of luminosity and
temperature. We find a minimum value of 0.08. This corresponds to a
mean deviation of $\sqrt{0.08}=28\%$ between the observations and the
model for wavelengths $<10~\rm{}\mu{}m$. Here, we require that good
fits not to exceed this mean deviation by more than a factor 2, i.e.,
we require $\chi_{\rm{}rel}^2<0.30\approx(2\times{}0.28)^2$. For certain
luminosity--temperature pairs, no solution exists, since no model
simultaneously fits the strict $J$--band and MIPS constraints.

The modeling constrains the internal luminosity to
\begin{displaymath}
L_{\rm int} = 0.08 ~ {\rm to} ~ 0.13 \, L_{\sun} \, .
\end{displaymath}
L1148--IRS does thus marginally fulfill the VeLLO luminosity criterion
of $L_{\rm int}\le{}0.1\,L_{\sun}$.  The luminosity is well
constrained, since it only requires a good SED match and does not
sensitively depend on model parameters. This is different for the
effective photospheric temperature, which we find to be
$T_{\rm{}eff}=2500~{\rm{}to}~5000~\rm{}K$. The temperature basically
follows from the SED slopes at wavelengths
$<10~\rm{}\mu{}m$. Unfortunately, at these wavelength the spherical
geometry of our model is probably too simplistic to capture the
relevant inner envelope extinction and scattering processes. This
induces further systematic uncertainties. In any event, though, a
central source of several $1000~\rm{}K$ must be present to explain the
detections in the 2MASS/FLAMINGOS bands.

The SED of L1148--IRS is thus consistent with the idea of L1148--IRS
being embedded in the dense core. In particular, this model provides a
quantitative explanation for the failure to detect the VeLLO with
SHARC--II. Initially, this lack of bright $350~\rm{}\mu{}m$ emission
was interpreted as evidence against L1148--IRS being an embedded VeLLO,
since all other VeLLOs show this emission
(\citealt{wu2007:c2d_sharc-ii}; \citealt{dunham2008:vellos}). Our SED
model shows, though, that no emission is to be expected, even if an
embedded source is present. Slightly more sensitive $350~\rm{}\mu{}m$
observations should be able to detect emission, though.

\section{VeLLO Natal Cores and Outflows}
Comprehensive understanding of L1148--IRS and other VeLLOs requires a
good knowledge of the environment in which these objects
form. Therefore, we launched an investigation of the properties of the
L1148 dense core. These results, as well as those for other VeLLOs,
are collected here to support our further discussion. We also include
additional data on outflows. Sections \ref{sec:props-cores} and
\ref{sec:props-outflows} summarize this information.\label{sec:catalogue}

\subsection{Data for L1148--IRS}
Section \ref{sec:previous-work} summarizes the data situation for
L1148. The information is collected from \citet{kauffmann2005:vellos}
and \citet{kauffmann2008:mambo-spitzer}. As described in this summary,
some of the data are also presented in Appendix
\ref{app-sec:dense-core}. Please refer to those sections for details.

\subsection{Data for other VeLLOs\label{sec:vello-cores}}
Here, we review the properties of known and proposed VeLLOs. First, we
assess the quality of VeLLO candidates. Specifically, we consider
objects that have previously been discussed to be similar to
L1014--IRS, have indications for luminosities $\le{}0.1\,L_{\sun}$, or
were discussed in the context of Brown Dwarf formation.

Visual inspection of Spitzer images of Oph B--11
\citep{greaves2003:oph_b11} does not show any far--infrared
counterpart. The core may actually be starless and is not considered
further here. We also tentatively remove Lupus 3 MMS
\citep{tachihara2007:Lupus-3} and L1415--IRS
\citep{stecklum2007:l1415-irs} from the VeLLO category. In these cases,
estimation of the internal luminosity on basis of the flux density at
$70~\rm{}\mu{}m$ wavelength \citep{dunham2008:vellos} suggests
luminosities significantly exceeding the VeLLO limit of
$0.1\,L_{\sun}$. We rate IC1396A:$\gamma_b$ \citep{reach2009:ic1396A}
as a source of unclear type, since the above luminosity estimate is
not possible for this source. Preliminary photometry of the VLA source
J025616+192703 in the L1457 dense core implies a low luminosity
\citep{heithausen2010:vello}, but data are too limited to judge the
probability that this really is a YSO and not a background object.
\citet{barrado2009:vello} present SSTB213 J041757, a possible VeLLO
candidate in Taurus. \citet{luhman2010:brown-dwarfs} show that this is
unlikely to be a Taurus YSO, though, and SSTB213 J041757 is thus
excluded here.

\citet{chen2010:first_core} and
\citet{enoch2010:first_core} have recently presented good candidates for
``first hydrostatic cores'' \citep{larson1969:numerical_collapse} with
internal luminosities $\lesssim{}0.01\,L_{\sun}$. These could be dense
cores which have just become optically thick to their own radiation,
before eventually collapsing into a more compact YSO.  Such objects
have a low age and luminosity, and they are embedded in a dense
core. This qualifies first hydrostatic cores as VeLLOs. The
\citeauthor{chen2010:first_core} and \citeauthor{enoch2010:first_core}
sources will be included in our comparison once their nature has been
established by decisive evidence.



Table \ref{tab:vello-properties} lists 7 well--studied VeLLOs in order
of decreasing aperture mass, $M_{4200\rm{}AU}$. It evaluates some
criteria for dense core evolution forwarded by
\citet{crapsi2005:survey}. By comparing properties like column
densities of dense gas tracers to characteristic thresholds, these
criteria essentially test whether a dense core contains a significant
mass of high density gas. Where appropriate (i.e., not when evaluating
the CO depletion factor: Appendix \ref{app-sec:abundances}), we
convert published values to match the dust properties adopted by
\citeauthor{kauffmann2008:mambo-spitzer}
(\citeyear{kauffmann2008:mambo-spitzer}; their Appendix A). We also
list internal luminosities and results of outflow searches, as
reported in the respective discovery paper. Distance--normalized flux
densities at $1.8~\rm{}cm$ wavelength, as reported by
\citet{scaife2010:ami-protostars}, characterize the radio emission
properties of the VeLLOs.

%
%
%

\begin{table*}
\begin{minipage}{\linewidth}
\setlength{\tabcolsep}{5pt}
\caption{VeLLOs and their Natal Cores\label{tab:vello-properties}}
\begin{tabular}{lllllllllllllllllll}
\hline
 & \multicolumn{3}{c}{Evolved Cores} & & \multicolumn{4}{c}{Unevolved Cores}\\ \cline{2-4} \cline{6-9}
Property & L1521F & Cha--MMS1 & IRAM04191 & & L673--7 & L1014 & L328 & L1148\\ \hline
\multicolumn{5}{l}{\emph{protostar properties:}}\\
outflow &
  Y & ? & Y & & Y & Y & ? & Y\\
extended outflow &
  n & n & Y & & Y & n & n & n\\
$L_{\rm int} / L_{\sun}$ &
  $\approx 0.04$ & [$\approx 0.01$] & $\approx 0.07$ & & $\approx 0.04$ & $\approx 0.09$ & $\approx 0.05$ & $\approx 0.10$\\
$F_{1.8\rm{}cm}d^2 / 10^{-2} ~ \rm mJy \, kpc^2$\vspace{1ex} & 0.3 & ? & 0.3 & & 0.5 & 1.9 & ? & $\le 0.7$ \\
\multicolumn{5}{l}{\emph{dense core properties:}}\\
$M_{4200 ~ \rm AU} / M_{\sun}$ &
  0.87 & [0.8] & 0.60 & & 0.39 & 0.36 & [$>0.13$/0.31] & 0.14\\
$\langle{}N({\rm{}H_2})\rangle_{4200~\rm{}AU} / 10^{22}\times{}{\rm{}cm^{-2}}$ &
 3.0 & [2.7] & 2.0 & & 1.3 & 1.2 & [$>0.4$/1.1] & 0.5\\
$N({\rm N_2D^+}) > 1.0 \times 10^{12} ~ \rm cm^{-2}$ &
  Y & [Y] & ? & & ? & n & n & n\\
$N({\rm N_2H^+}) > 8.5 \times 10^{12} ~ \rm cm^{-2}$ &
  Y & [Y] & Y & & ? & n & Y & n\\
$N({\rm N_2D^+}) / N({\rm N_2H^+}) \ge 0.1$ &
  Y & Y & ? & & ? & Y & n & ?\\
$f_{\rm d}({\rm CO}) > 10$ &
  Y & ? & ? & & ? & n & n & n\\
$n_{\rm c}({\rm H_2}) > 2.5 \times 10^5 ~ \rm cm^{-3}$ &
  Y & Y & Y & & n & n & n & n\\
$\Delta v ({\rm N_2H^+}) > 0.25 ~ \rm km ~ s^{-1}$ &
  Y & Y & Y & & Y & Y & Y & n\\
inward motions&
  Y & ? & Y & & n & n & n & Y\\
$r_{70\%} < 4 \, 800 ~ \rm AU$ &
  Y & Y & Y & & Y & Y & Y & Y\\
\hline
\end{tabular}

\vspace{0.5ex} \textbf{References.}  Data are from
\citet{lee1999:contr_survey, lee2009:l328, andre1999:iram04191,
  motte2001:protostars, belloche2002:iram04191, belloche2006:cha-mm1,
  young2004:l1014, crapsi2005:survey, crapsi2005:l1014,
  bourke2005:l1014-outflow, bourke2006:l1521f, dunham2006:iram_04191,
  dunham2010:l673-7}; \citeauthor{kauffmann2005:vellos}
(\citeyear{kauffmann2005:vellos},
\citeyear{kauffmann2008:mambo-spitzer}, and this work); and
\citet{scaife2010:ami-protostars}\smallskip

\textbf{Notes.}  The dense core properties are compared to the
criteria for ``evolved'' cores by \citet{crapsi2005:survey} after
conversion of the limiting density to our choice of dust emission
properties. A `Y' marker indicates that a criterion is met, a `n'
marker that it is not, a question mark that the data are insufficient
to decide. Properties marked by square brackets ([\dots]) are not
readily available from the relevant publications and must be gleaned
from the other published properties. For Cha--MMS1, the $\rm{}N_2H^+$
and $\rm{}N_2D^+$ column densities have to be derived from the
abundances and dust column densities reported by
\citet{belloche2006:cha-mm1}. The aperture mass is derived from models
for the density structure presented in the same publication. For
L328--IRS, the envelope masses are from lower limits of the
$1.2~\rm{}mm$ wavelength dust emission (\citealt{lee2009:l328};
$70~\rm{}mJy$ within $2000~\rm{}AU$ radius) and integration of the
\citeauthor{lee2009:l328} envelope density model to $4200~\rm{}AU$
radius. We deviate from the \citeauthor{crapsi2005:survey} practice to
infer inward motions from asymmetries in the $\rm{}N_2H^+$ lines;
instead, inward motions are either inferred from the analysis of line
shifts (L1521F, L673--7, L1014, L328, and L1148), or from detailed
modeling of line shapes (IRAM04191). For all sources the uncertainty
in the internal luminosity is of order $\pm{}0.03\,L_{\sun}$.
\end{minipage}
\end{table*}

\subsection{Properties of VeLLO Natal Cores\label{sec:props-cores}}
The VeLLOs in Table \ref{tab:vello-properties} can roughly
be divided into two groups. The first group is composed of L1521F,
Cha--MMS1, and IRAM04191. These cores fail in none of the
\citet{crapsi2005:survey} criteria for ``evolved'' dense cores (if we
stick only to criteria that can be tested with existing data). For
instance, these cores are very dense (in terms of volume and column
density), they have strongly peaked density profile (as indicated by
small values of $r_{70\%}$, i.e.,the radius of the dust emission
contour at 70\% peak intensity; \citealt{crapsi2005:survey}), and the
inner regions of the dense core are in an advanced stage of chemical
evolution (i.e., CO depletion, and production of deuterated
$\rm{}N_2H^+$). \citet{crapsi2005:survey} argued that their criteria
select dense cores relatively close to the onset of protostellar
collapse. It is thus not surprising to find that collapse has indeed
started in some of these objects.

The second group is composed of ``unevolved'' dense cores, that fail
in at least one of the \citeauthor{crapsi2005:survey} criteria; these
cores have relatively low (column) densities, and there are often no
signs of advanced chemistry. Following \citet{crapsi2005:survey},
these dense cores are not expected to collapse and form stars. Section
\ref{sec:l1148-formation} explores how stars might still have formed
in these cores. The most prominent member of this group is L1014; the
discovery of a Spitzer source in this core of low density, which was
believed to be ``starless'', was a big surprise
\citep{young2004:l1014}. Also L1148 belongs to this group.

\emph{The terminology used above is not meant to indicate that
  ``unevolved'' cores mature and eventually become ``evolved'' ones!}
It is only used to indicate the relation to the
\citeauthor{crapsi2005:survey} criteria.\medskip

\noindent{}In summary, some VeLLOs fulfil the
\citet{crapsi2005:survey} criteria for evolved cores, while others do
not. Since star formation is ongoing in all cores listed in Table
\ref{tab:vello-properties}, the initial conditions for stars to form
appear to be very diverse.

L1148--IRS different from the other VeLLOs in that this source appears
to have the lowest aperture mass of all VeLLOs. It also features the
lowest $\rm{}N_2H^+$ line width ($[0.232\pm{}0.048]~\rm{}km\,s^{-1}$;
the next higher value for VeLLOs is $0.3~\rm{}km\,s^{-1}$ for
L1521F). Finally, L1148 is the only ``unevolved'' VeLLO core in which
inward motions have been detected. Since line asymmetries are easier
detected for narrow lines (Eq.\ \ref{eq-app:velocity-differences}),
this might be a consequence of the low $\rm{}N_2H^+$ line width.

\subsection{Properties of VeLLO Outflows\label{sec:props-outflows}}
VeLLOs differ vastly in their outflow properties. For example, only 2
out of the 7 VeLLOs discussed in Table \ref{tab:vello-properties}
(i.e., IRAM04191 and L673--7--IRS) drive a ``prominent'' extended outflow
that can easily be detected using single--dish telescopes. 
The remaining VeLLOs all appear to drive outflows (based on Spitzer
image morphology and interferometer observations), but sensitive
high--resolution observations are needed to detect these (like those
presented in Section \ref{sec:outflow}).

As seen in Table \ref{tab:vello-properties}, the presence of prominent
outflow activity does not appear to correlate with any other dense
core property. In the present sample, to be specific, an extended
outflow is driven by a VeLLO in one of the most ``evolved'' cores,
i.e.\ IRAM04191, which also has a large aperture mass. But also the
VeLLO in L673--7 has a prominent outflow, despite a very moderate
aperture mass and failure in some of the
\citeauthor{crapsi2005:survey} criteria. In addition, cores more
evolved than IRAM04191 (e.g., L1521F), as well as cores very similar
to L673--7 (e.g., L1014), do not drive such flow. This diversity
could be a consequence of VeLLO evolution; VeLLOs without outflows
might be in a stage preceding or following an accretion phase, in
which prominent outflows are produced. It is not clear whether other
options --- i.e., VeLLO formation without prominent outflows in any
evolutionary stage --- provide credible models.

It is also not clear how the presence of prominent outflows relates to
the concept of ``episodic accretion'', i.e., accretion (and thus
outflow generation) occurring only during short episodes of the
protostellar lifetime \citep{kenyon1990:iras-survey}. In this picture,
VeLLOs would quasi--periodically create and gradually loose their
prominent outflows. Dynamical ages for the outflows of IRAM04191 and
L673--7--IRS suggest that individual outflow periods\footnote{Prominent
  outflows are detectable during this period. This does, however, not
  require that they are continuously driven during this time.} last a
few $10^4~\rm{}yr$ in VeLLOs. If VeLLOs without prominent outflows are
indeed in a dormant stage, this stage should prevail for an even
longer timescale ($>10^4~\rm{}yr$), as indicated by the large fraction
of VeLLOs without such outflows. Combination of Spitzer source counts
\citep{evans2009:c2d-summary} with evolutionary models
\citep{dunham2010:evol-indicators} indeed shows that YSOs might
accrete most of their mass in $\lesssim{}10$ bursts occurring over a
period $\sim{}2\times{}10^5~\rm{}yr$. These models do thus imply
dormant periods of several $10^4~\rm{}yr$, just as suggested by VeLLO
outflow statistics.

\section{Discussion: L1148--IRS in Context}
\label{sec:discussion}

\subsection{Evolutionary Stage \& Mass of
  L1148--IRS\label{sec:stage-mass}}
L1148--IRS has a bolometric temperature $(110\pm{}15)~\rm{}K$ (Section
\ref{sec:sed-observations}). This implies that L11148--IRS belongs to
the protostellar class I, and suggests an age $<7\times{}10^5~\rm{}yr$
\citep{evans2009:c2d-summary}. We use this limit in the
following.\medskip

\noindent{}As shown by \citet{young2004:l1014}, the internal
luminosity of $\le{}0.1\,L_{\sun}$ limits the present mass of all
VeLLOs to a most likely value $\lesssim{}0.1\,M_{\sun}$, if the
luminosity is dominated by accretion power. This is probably the case
for L1148--IRS, since it appears to drive an (supposedly
accretion--powered) outflow. Further, the young brown dwarf FU~Tau (a
class II object of $\lesssim{}10^6~\rm{}yr$) has a luminosity of
$0.2\,L_{\sun}$ (\citealt{luhman2009:fu-tau}; $\ge{}0.1\,L_{\sun}$ for
the brightest component, if FU~Tau is an unresolved binary). This
object might be unusually luminous \citep{luhman2009:fu-tau},
but its properties support the notion that YSO luminosities
$\lesssim{}0.1\,L_{\sun}$ imply a mass in the brown dwarf
domain. Thus, we find that the most likely present--day mass is
$\lesssim{}0.1\,M_{\sun}$.

Comparison with evolutionary tracks by \citet{baraffe1998:tracks} and
\citet{chabrier2000:tracks} suggest a very conservative mass limit
$\ll{}0.35\,M_{\sun}$ (when reducing the model luminosities by a
factor 3: \citealt{reiners2005:calibration}). Models by
\citet{wuchterl2003:evolution} suggest $<0.1\,M_{\sun}$, if one only
considers early stages with significant accretion. Unfortunately,
their models are not well tested against observations.\medskip

\noindent{}To understand the collapse conditions in L1148, consider
the collapse of a hydrostatic isothermal sphere of radius $R$ that is
subject to an environmental pressure, $P_{\rm{}env}$. To be unstable,
the sphere must have a mass
\begin{equation}
M > 2.0 \, M_{\sun} \,
\left( \frac{T_{\rm g}}{10 ~ \rm K} \right)^2 \,
\left( \frac{R}{0.1 ~ \rm pc} \right) \, ,
\label{eq:m_cr-radius}
\end{equation}
and the pressure must obey
\begin{equation}
\frac{P_{\rm ext}}{k_{\rm B}} > 3.3 \times 10^4 \, {\rm cm^{-3}} \,
\left( \frac{T_{\rm g}}{10 ~ \rm K} \right)^4 \,
\left( \frac{R}{0.1 ~ \rm pc} \right)^{-2}
\label{eq:p_cr}
\end{equation}
(\citealt{bonnor1956:be-spheres} and \citealt{ebert1955:be-spheres};
$k_{\rm{}B}$ is Boltzmann's constant). Star formation requires that
the inner parts of the protostellar envelope are unstable\footnote{To
  be specific, this is a requirement only for models with inside--out
  collapse. The evolution of these collapsing spheres is similar to
  those studied by \citet{shu1977:self-sim_collapse}. One major
  difference, though, is that \citeauthor{shu1977:self-sim_collapse}
  considers --- strictly speaking --- infinite spheres with singular
  density profiles. His structures are unstable to perturbations at
  all radii, just as required here.}. We can use the aforementioned
aperture mass, $M_{4200\rm{}AU}$, to test this condition. For
$R=4200~\rm{}AU$ and $T_{\rm{}g}=10~\rm{}K$, Eq.\
(\ref{eq:m_cr-radius}) yields the condition $M>0.41\,M_{\sun}$. Thus,
$M_{4200\rm{}AU}$ fails the condition for gravitational
instability. The reservoir actively forming the VeLLO must therefore
have a radius $<4200~\rm{}AU$. This implies an unstable mass smaller
$M_{4200\rm{}AU}$, i.e.\ $<0.14\,M_{\sun}$ (Section
\ref{sec:previous-work}). Radii $<4200~\rm{}AU$ also imply a pressure
$P_{\rm{}ext}/k_{\rm{}B}>8\times{}10^5~\rm{}cm^{-2}$. As we discuss in
Section \ref{sec:l1148-formation}, this pressure is relatively
high. This has implications on VeLLO formation models.

%
%

Combined with the present--day mass of L1148--IRS, $<0.1\,M_{\sun}$,
this estimate of the collapsing mass suggests a final mass
$<0.24\,M_{\sun}$. If this VeLLO has a present--day mass
$<0.05\,M_{\sun}$, and the immediate envelope of $<0.14\,M_{\sun}$ is
accreted with an efficiency $\sim{}1/3$ (as deemed typical; e.g.,
\citealt{alves2007:imf}), then L1148--IRS could have a final mass even
below the brown dwarf limit. After L328--IRS \citep{lee2009:l328}, this
is the second VeLLO for which such a small final mass is derived.

\subsection{The Nature and Evolution of
  VeLLOs\label{sec:vello-nature}}
There is probably general consent that VeLLOs have a most likely
present--day mass $<0.1\,M_{\sun}$ (Section
\ref{sec:stage-mass}). The final mass at the end of the accretion
phase might, however, be much larger.

VeLLO natal cores differ significantly in the mass of gas at high
density they contain (Section \ref{sec:catalogue}). Since VeLLOs
form from this gas, these differences may suggest that different
VeLLOs take different evolutionary paths
\citep{bourke2006:l1521f}. Some may have significant future accretion
and attain a stellar final mass \citep{andre1999:iram04191}, while
lack of accretion in others could render these objects young brown
dwarfs \citep{young2004:l1014, huard2006:l1014}. Table
\ref{tab:vello-properties} thus presents a first attempt to
systematically characterize the likeliness of these different
evolutionary paths for several VeLLOs. The number of positive
indicators for significant amounts of high density gas near the VeLLO
increase towards the left of Table \ref{tab:vello-properties}. The
VeLLO final mass is therefore likely to increase in this direction
too. L328--IRS and L1148--IRS hold a peculiar position among VeLLOs,
given that their most likely final mass is very low ($<0.15\,M_{\sun}$
and $<0.24\,M_{\sun}$, respectively; see \citealt{lee2009:l328} and
Section \ref{sec:stage-mass}).

\subsection{Formation of Brown Dwarfs: Lessons from
  L1148\label{sec:l1148-formation}}
One unsolved aspect of isolated brown dwarf (BD) formation is the
large difference between the final BD mass,
$M_{\rm{}BD}\lesssim0.08\,M_{\sun}$, and the expected mass of the
collapsing natal dense core, $M$: in most cases
$M\gg{}M_{\rm{}BD}$. L1148--IRS can be used to study this problem, even
in case this VeLLO has a final mass exceeding the BD mass
limit.\medskip

\noindent{}Consider once again pressure--bound hydrostatic isothermal
spheres. Combination of Eqs.\ (\ref{eq:m_cr-radius}, \ref{eq:p_cr})
yields the condition
\begin{equation}
M > 1.2 \, M_{\sun} \,
\left(\frac{T_{\rm g}}{10 ~ \rm K}\right)^4 \,
\left(\frac{P_{\rm env} / k_{\rm B}}{10^5 ~ \rm K \, cm^{-3}}\right)^{-1/2}
\label{eq:m-jeans}
\end{equation}
for unstable spheres. This system will form a star (or BD) of mass
$M_{\star}=\eta\,M$, where $\eta\le{}1$ is the star formation
efficiency. For typical temperatures ($\approx{}10~\rm{}K$) and
pressures (a few $10^5~\rm{}K\,cm^{-3}$; e.g.\
\citealt{bertoldi1992:pr_conf_cores}), one obtains the collapse
condition $M\gtrsim1\,M_{\sun}$. Typical values of $\eta$ are believed
to be $\sim1/3$ (e.g., \citealt{alves2007:imf}), yielding typical
stellar masses $\sim{}0.3\,M_{\sun}$.

In this framework, to achieve lower stellar masses, BD formation
requires unusually small temperature ($T_{\rm{}g}\ll{}10~\rm{}K$),
large pressure ($P_{\rm{}env}/k_{\rm{}B}\gg{}10^5~\rm{}K\,cm^{-3}$),
low efficiency ($\eta\ll{}1/3$), or a combination of these. L1148--IRS
provides a test case for such reasoning: Eq.\ (\ref{eq:m_cr-radius})
suggests $M<0.14\,M_{\sun}$ for the collapsing core in this region.

The structure of the L1148 core thus appears to require peculiar
environmental conditions. We have no satisfying explanation why and in
which sense (temperature, pressure, etc.)  these should prevail in
L1148. One can, however, speculate that the unusual kinematic
conditions in this region might have played a role: multiple velocity
components (in $\rm C^{18}O$), velocity jumps near L1148--IRS (in CCS)
and contraction motions on spatial scales $\gtrsim{}0.3~\rm{}pc$ are
not commonly observed in dense cores (see Section
\ref{sec:previous-work} and Appendix \ref{app-sec:velocity-field}).

The environmental conditions in L1148 have actually been explored by
\citet{nutter2009:cepheus-flare}. Spatial offsets between intensity
peaks at different wavelengths are consistent with external heating of
the L1148 core. One could imagine that the core is sculpted by such a
radiation field. \citet{kirk2009:cepheus-flare} 
suggests indeed that the entire Cepheus Flare Complex is shaped
by external triggering. Related externally triggered compression
events would be suited to produce low critical masses via an increase
in $P_{\rm{}env}$.\medskip

\noindent{}It might, of course, also be that the initial unstable mass
of the dense core was $\sim{}1\,M_{\sun}$, as suggested by Eq.\
(\ref{eq:m-jeans}), but that we now observe a much more evolved phase,
where the initial mass is no longer relevant. \citet{lee2009:l328},
for instance, suggest that L328--IRS might have moved out of its natal
core. This would leave this VeLLO with a small mass reservoir to
accrete from. \citet{huard2006:l1014} present a similar argument for
L1014--IRS. The complex kinematics in L1148 might suggest such a
scenario also for L1148--IRS. In fact, Section \ref{sec:previous-work}
shows that there is a possible --- though not significant ---
$7\farcs{}6$ spatial offset (i.e., $2500~\rm{}AU$) between L1148--IRS
and the core's column density peak. The growth of L1148--IRS might
thus be stinted, because it has detached from the dense core from
which it cannot accrete from it any longer.

Similarly, L1148--IRS could be in a very advanced stage of evolution
where most of the envelope has been dispersed. In principle, strong
VeLLO outflows (e.g., as those from IRAM04191;
\citealt{andre1999:iram04191}) might be able to do this.

The credibility of these scenarios is unclear, though. Consider
L1148--IRS as an example. Section \ref{sec:stage-mass} suggests that
the present--day mass of this VeLLO is $\lesssim{}0.1\,M_{\sun}$. The
immediate environment of L1148--IRS contains $\sim{}0.14\,M_{\sun}$
(from the $4200~\rm{}AU$ aperture mass). If outflows have dispersed
the gas, they must thus have disrupted a core of about
$0.75\,M_{\sun}$. It is not known whether dispersal by outflows is
efficient enough to do this \citep{arce2007:outflows}. Similarly, it
is not obvious how VeLLOs should be able to leave their parental dense
core. That would require that accretion from the gas reservoir onto
the VeLLO increases the relative velocity between the gas reservoir
and the VeLLO. This seems unlikely.

\section{Summary}
We present a detailed analysis of a candidate VeLLO in the Cepheus
Flare, L1148--IRS. This work goes beyond the initial announcement of
discovery \citep{kauffmann2005:vellos} by presenting: (\textit{i}) a
comprehensive and sensitive search for a CO outflow; (\textit{ii})
detailed modelling of the observed spectral energy distribution; and
(\textit{iii}) extensive discussions of the properties of the natal
dense core. In addition, (\textit{iv}) the analysis is augmented by
presenting L1148--IRS in the context of an up--to--date VeLLO
catalogue. The main conclusions are as
follows.\label{sec:summary}\medskip

\noindent{}(1) L1148--IRS drives a compact ($\sim{}5\arcsec$, about
$1800~\rm{}AU$ in projection) CO outflow (Section
\ref{sec:outflow}). This confirms L1148--IRS as a clear--cut protostar,
and associates it with L1148 through its velocity. The outflow can
only be detected in interferometer maps, such as those from the
Plateau de Bure Interferometer presented here; single--dish data, e.g.\
from the IRAM 30m--telescope, do not reveal it. Weak compact
($\sim{}2\arcsec$) probable outflow nebulosity is seen near $2\micron$
wavelength.\smallskip

\noindent{}(2) Modelling of the spectral energy distribution suggests
the presence of a point source of $0.08~{\rm{}to}~0.13\,{}L_{\sun}$
and $T_{\rm{}eff}=2500~{\rm{}to}~5000~\rm{}K$ effective photospheric
temperature (Section \ref{sec:sed}). Strong emission near $100\micron$
wavelength requires this source to be deeply embedded in the natal
dense core; it corresponds to a source of infrared class 0 or
I. L1148--IRS does therefore meet all criteria to be a VeLLO (i.e.,
$L\le{}0.1\,{}L_{\sun}$ and embeddedness in a core), though this is
just marginally the case for the luminosity.\smallskip

\noindent{}(3) The natal dense core is characterized by densities and
column densities that are unusually low for a star--forming core
(Section \ref{sec:previous-work}). Multiple velocity components,
velocity jumps, and velocity gradients are detected in the core,
pointing towards unusual kinematic conditions in this region. Strong
evidence for contraction motions is found in asymmetries of optically
thick lines.\smallskip

\noindent{}(4) Among all VeLLOs known, L1148--IRS stands out as the one
with the lowest envelope mass and a core of unusually low (column)
density (Table \ref{tab:vello-properties} in Section
\ref{sec:discussion}). It is a riddle how a star could have formed
under these conditions.\medskip

\noindent{}As in all other VeLLOs, the low luminosity suggests a substellar
present mass for L1148--IRS, i.e.\ $< 0.1 \, M_{\sun}$. The
gravitationally unstable mass surrounding L1148--IRS appears to be
$\lesssim 0.14 \, M_{\sun}$. Their sum suggests a very low final mass,
so that L1148--IRS may be a good candidate for a proto brown dwarf
forming in isolation.

\section*{Acknowledgements}
We are indebted to our local IRAM contacts, Arancha Castro--Carrizo
and Fr\'ed\'eric Gueth, who changed their busy schedule to simplify
our travel arrangements for a data reduction visit. We thank Endrik
Kr\"ugel for providing us with a custom--modified dust radiative
transfer code. Mike Dunham was so kind to check our VeLLO catalogue
(Table \ref{tab:vello-properties}, Section \ref{sec:vello-cores})
against his notes. JK is deeply indebted to FB, Karl Menten, Malcolm
Walmsley, Johannes Schmid--Burgk, and TP. This work would have been
impossible without their continuous and unconditional support. He also
thanks Di Li and Paul Goldsmith, his hosts at JPL, for making this
research possible. This project was supported by an appointment of JK
to the NASA Postdoctoral Program at the Jet Propulsion Laboratory,
administered by Oak Ridge Associated Universities through a contract
with NASA. His research was executed at the Jet Propulsion Laboratory,
California Institute of Technology, under a contract with the National
Air and Space Administration. Partial support for TLB was provided by
NASA through contracts 1279198 and 1288806 issued by the Jet
Propulsion Laboratory, California Institute of Technology, under NASA
contract 1407. CWL acknowledges the support by Basic Science Research
Program through the National Research Foundation of Korea funded by
the Ministry of Education, Science and Technology (2010--0011605).

\bibliographystyle{apj}
\bibliography{mendeley/library}

\appendix

\section{Photometry for L1148--IRS}
Table \ref{tab-l1148:L1148_SED} presents the SED of L1148--IRS,
compiled from data sources listed in Section
\ref{sec:sed-observations}. Here we give some further processing
details.\label{app:photometry}

In the IRAC and the MIPS1 band, we average results, if more than one
flux density measurement is available. For the MIPS2 and MIPS3 band,
we use our own aperture photometry results instead of the values
derived by point--source fitting; this is likely to be better suited
for the slightly extended emission. The SHARC--II and SCUBA upper
limits are derived by adding the observed flux and the noise level
times three (our SHARC--II limits supersede those from
\citealt{wu2007:c2d_sharc-ii}). We check our photometry against
further Spitzer MIPS \citep{kirk2009:cepheus-flare} and independent
Akari \citep{nutter2009:cepheus-flare} studies of the
region\footnote{D.\ Nutter provided us with dedicated photometry,
  based on the data from his Akari study
  \citep{nutter2009:cepheus-flare}. We are grateful to him and J.\
  Kirk for providing us their results in advance of publication.}. The
data are usually consistent within their uncertainties. A notable
exception is the Akari flux of $130 ~ \rm mJy$ at $90 ~ \rm \mu m$
wavelength; this is inconsistent with the much higher MIPS flux
densities at 70 and $160\micron$. We have no explanation for this, but
elect to rely on our higher quality Spitzer data.


\begin{table*}
\begin{minipage}{\linewidth}
\caption{The spectral energy distribution of
  L1148--IRS\label{tab-l1148:L1148_SED}}
\begin{tabular}{llllllll}
\hline
Band & Wave Length & Bandwidth & Resolution & Apertures &
Flux Density & Calibration Uncertainty\\
& $\mathrm{\mu m}$ & $\mathrm{\mu m}$ & arcsec & arcsec &
mJy\\ \hline
2MASS $J$ & 1.24 & 0.16 & 2.8 & PSF & $< 0.1$ & 2\%\\
FLAMINGOS $J$ & 1.24 & 0.16 & 1.3 & 2.56, 6.4 -- 12.8 & $< 0.1$ & 2\%\\
2MASS $H$ & 1.66 & 0.25 & 2.8 & PSF & $< 0.6$ & 2\%\\
FLAMINGOS $H$ & 1.66 & 0.25 & 1.3 & 2.56, 6.4 -- 12.8 &
  $0.421 \pm 0.021$ & 2\%\\
2MASS $K_{\mathrm{s}}$ & 2.16 & 0.26 & 2.8 & PSF & 
  $0.9 \pm 0.1$ & 2\%\\
FLAMINGOS $K_{\mathrm{s}}$ & 2.16 & 0.26 & 1.3 & 2.56, 6.4 -- 12.8 &
  $1.304 \pm 0.065$ & 2\%\\
IRAC 1 & 3.6 & 0.8 & 1.7 & PSF & $1.22$ & 15\%\\
IRAC 2 & 4.5 & 1.0 & 1.7 & PSF & $1.33$ & 15\%\\
IRAC 3 & 5.7 & 1.4 & 1.9 & PSF & $0.83$ & 15\%\\
IRAC 4 & 7.9 & 2.9 & 2.0 & PSF & $1.14$ & 15\%\\
IRAS 12 & 12 & 7 & 25 & PSF & $< 77$ & 3\%\\
MIPS 1 & 24 & 5 & 6 & PSF & $72.5$ & 15\%\\
IRAS 25 & 25 & 11 & 25 & PSF & $156$ & 6\%\\
IRAS 60 & 60 & 33 & 60 & PSF & $283$ & 6\%\\
MIPS 2 & 71 & 19 & 18 & 60, 80 -- 120 & $326$ & 20\%\\
IRAS 100 & 100 & 31.5 & 100 & PSF & $< 3955$ & 10\%\\
MIPS 3 & 156 & 35 & 40 & 48, 128 -- 256 & $\approx 500$ & 50\%\\
SHARC--II & 350 & 45 & 9 & 25.8 & $< 608$ & 15\%\\
SCUBA & 850 & 170 & 14 & 22.9 & $< 159$ & 20\%\\
MAMBO & 1200 & 210 & 11 & 25.8 & $33$ & 20\%\\ \hline
\end{tabular}

\vspace{0.5ex} \textbf{Notes.}  For every band the central wavelength
and width of the filter, the telescope resolution, the apertures used
for photometry, the measured flux density including color corrections
(where necessary) and its uncertainty due to noise, and the
calibration uncertainty are given. For most bands photometry is done
by point--source fitting. Then the aperture to that the flux density
refers is given by the point spread function (PSF) and we mark these
bands by listing ``PSF'' as the used aperture. For the other bands,
where we perform aperture photometry, the first number given for the
aperture is the diameter across which the intensity is integrated,
while the given range indicates the diameter of the inner and outer
circle enclosing the area used to obtain an estimate for the
background intensity (does not apply to bolometer data). The
uncertainty in the MIPS 3 band reflects the flux density variation due
to different aperture choices.
\end{minipage}
\end{table*}

\section{The L1148 Dense Core}
As already explained in Section \ref{sec:previous-work},
\citet{kauffmann2005:vellos} lacked the space to describe their data
in detail. To provide a proper reference, we do therefore present
those observations in this appendix. We also discuss new observations
of contraction motions in L1148 (Appendix
\ref{app-sec:contraction}). To improve the readability of this paper,
their rather straightforward discussion is removed from the main
text.\label{app-sec:dense-core}

In this appendix, we do therefore first introduce a previously rather
undocumented data set that had never been documented well, i.e., our
Effelsberg observations (Appendix \ref{app-sec:effelsberg}). We then
discuss the spatial mass distribution of the core in more detail than
appropriate in the main text (Appendix
\ref{app-sec:mass-distribution}). This is followed by extended
discussions of the line emission data first used by
\citet{kauffmann2005:vellos} (Appendix \ref{app-sec:velocity-field}:
velocity field of the cloud; Appendix \ref{app-sec:abundances}:
molecular abundances). We finish with an analysis of new data on
contraction motions in the cloud (Appendix
\ref{app-sec:contraction}).

\subsection{Additional Data: Effelsberg
  100m--telescope\label{app-sec:effelsberg}}
Spectroscopic observations with the Effelsberg 100m--telescope
(operated by the Max--Planck--Institut f\"ur Radioastronomie in Bonn,
Germany) were done 2005 April 13 and April 29 to May 2. The
($J,F=2,1$--$1,0$) transition of CCS was observed with the facility
$1.3~\rm{}cm$ primary focus spectral line receiver and the AK90
autocorrelator as the backend. All data was taken in the frequency
switching mode. The pointing corrections in azimuth and elevation were
mostly below $6\arcsec$, but corrections of $15\arcsec$ occurred
once. The focus corrections were below $0.6~\rm{}mm$. Pointing was
checked at least once in 4 hours and the focus corrected at least once
in 6 hours. Further details are listed in Table
\ref{app-tab-l1148:radio_obs}.

The spectroscopic observations were reduced using CLASS. We removed
low--order baselines and partially used Fourier (FFT) filtering to
further improve the baseline quality. Further details of the data
reduction are given in the respective sections below.

The data were calibrated using cross--scans towards 3C48, 3C123,
3C147, 3C286, and NGC7027. The observed telescope response is combined
with the flux densities given by \citet{ott1994:calibrators} to
determine the calibration factor. We use a factor of 3.62 to convert
the dimensionless telescope counts into main beam brightness
temperature. Please refer to
\path|http://www.mpifr-bonn.mpg.de/staff/tpillai/eff_calib/eff_calib.html|
for more details.

%
%
%
%

\begin{table*}
\begin{minipage}{\linewidth}
\caption{Spectroscopic observations at the IRAM 30m--telescope and the
  Effelsberg 100m--telescope\label{app-tab-l1148:radio_obs}}
\begin{tabular}{lllllllll}
\hline
Molecule & Transition & Frequency & ${T_{\rm{}mb}}^{\rm{}a}$ &
  Vel.\ Uncert.$^{\rm b}$ & Beam Size & Freq.\ Throw$^{\rm c}$ &
  $T_{\rm sys}${}$^{\rm d}$ & Vel.\ Resol.$^{\rm e}$\vspace{1ex}\\
& & GHz & K & $\mathrm{km ~ s^{-1}}$ & arcsec &
  MHz & K & $\mathrm{km ~ s^{-1}}$\\ \hline
\multicolumn{4}{l}{\itshape Effelsberg:}\\
$\mathrm{CCS}$ & $J,F = 2, 1$--$1,0$ & 22.3440308$^1$ & $1.43\pm{}0.32$ &
  0.013 & 39 & 2.5 & 271 & 0.033\\
\\
\multicolumn{4}{l}{\itshape IRAM 30m-telescope:}\\
$\mathrm{N_2H^+}$ & $J,F_1,F =$ & 93.1762608$^2$ & $0.27\pm{}0.05$ &
  0.019 & 26 & 7.0 & 142$^{\rm f}$ & 0.063$^{\rm f}$\\
  & \quad $1,0,1$--$0,1,2$ & & & & & & 196$^{\rm g}$ & 0.021$^{\rm g}$\\
$\mathrm{N_2D^+}$ & $J,F_1,F =$ & 77.1121180$^2$ & $<0.21$ &
  0.023 & 31 & 7.8 & 267 & 0.076\\
  & \quad $1,0,1$--$0,1,2$\\
$\mathrm{CS}$ & $J = 2$--1 & 97.980953$^3$ & $0.30\pm{}0.04$ & 0.006 & 25 &
  7.0 & 145 & 0.020\\
  & $J = 3$--2 & 146.969029$^3$ & $<0.22$ & 0.002 & 16 & 5.0 & 363 & 0.040\\
$\mathrm{^{12}CO}$ & $J = 2$--1 & 230.538000$^4$ & $4.87\pm{}0.14$ & 0.001 & 10
  & 15 & $377^{\rm f}$ & $0.051^{\rm f}$\\
$\mathrm{C^{18}O}$ & $J = 2$--1 & 219.560358$^5$ & $1.00\pm{}0.10$ & 0.001 & 11 &
 15 & 291 & 0.053\\
$\mathrm{HCO^+}$ & $J = 1$--0 & 89.1884957$^1$ & $0.64\pm{}0.04$ & 0.037 & 27 &
  7.0 & 132 & 0.023\\
$\mathrm{H^{13}CO^+}$ & $J = 1$--0 ``primary'' & 86.7542982$^6$ &
$0.20\pm{}0.04$ & 0.014 & 28 & 7.0 & 111 & 0.045\\
$\mathrm{HCN}$ & $J,F = 1,1$--$0,1$ & 88.6304157$^1$ & $0.05\pm{}0.02$ & 0.003 & 27 &
  7.0 & 131 & 0.066\\
  & $J,F = 1,2$--$0,1$ & 88.6318473$^1$ & $0.16\pm{}0.02$ &
  \multicolumn{5}{c}{\dotfill ditto \dotfill}\\
  & $J,F = 1,0$--$0,1$ & 88.6339360$^1$ & $0.09\pm{}0.02$ &
  \multicolumn{5}{c}{\dotfill ditto \dotfill}\\
\hline
\end{tabular}

\vspace{0.5ex} \textbf{References.}
(1) \citet{mueller2005:cdms}; (2) \citet{dore2004:n2d+}; (3)
\citet{gottlieb2003:cs}; (4) \citet{winnewisser1997:co}; (5)
\citet{cazzoli2003:c18o}; (6) \citet{schmid-burgk2004:l1512}\smallskip

\textbf{Notes.}
$^{\rm a}$ peak main beam brightness temperature per spectrum (and
noise level measured as the root mean square deviation at velocities free
of line emission) at the
position closest to L1148--IRS, determined using Gaussian fits to the
lines, where dashes (``--'') indicate non--detections, and upper
limits are the noise level times three;
$^{\rm b}$ uncertainty in frequency, and implied uncertainty in
velocity;
$^{\rm c}$ frequency throw in frequency--switched observations;
$^{\rm d}$ system temperature, in main beam brightness scale;
$^{\rm e}$ velocity resolution of the spectral observations, after
smoothing if applicable;
$^{\rm f}$ for mapping observations in December 2004;
$^{\rm g}$ for pointed observations in July 2005
\end{minipage}
\end{table*}

\subsection{Spatial Mass
  Distribution\label{app-sec:mass-distribution}}
Based on the dust emission maps shown in Figures \ref{fig:discovery}
and \ref{fig:layout-observations}, \citet{kauffmann2008:mambo-spitzer}
noted that the L1148 region contains two parallel filaments of
$3\,M_{\sun}$ and $9\,M_{\sun}$, respectively of
$1\arcmin{}\times{}5\arcmin$ size (i.e, $0.1\times{}0.5~\rm{}pc^2$;
both filaments have a similar size). The north--western filament
contains L1148--IRS and has the lower mass. The $\rm H_2$ peak column
densities are $6\times{}10^{21}~\rm{}cm^{-2}$ and
$8\times{}10^{21}~\rm{}cm^{-2}$.

To derive radial density laws, we execute a new analysis of cuts
through the dust emission peak near L1148--IRS. They are taken at a
position angle $135\deg$, i.e., perpendicular to the filament
axis. The cuts (i.e., dust emission intensity as a function of offset
from the peak) are compared to intensity profiles expected for density
laws $\varrho(r)\propto^{-\alpha}$ and constant dust temperature. The
synthetic intensities are calculated taking the limited telescope
resolution into account. Within the uncertainties, the dust emission
distribution is consistent with pure power--laws with
$\alpha=1.75~{\rm{}to}~2.0$. In other words, compared to the beam
diameter, regions of constant density (i.e., $\alpha=0$) cannot have a
significant size.

\subsection{Velocity Field\label{app-sec:velocity-field}}
\citet{kauffmann2005:vellos} used CCS ($J_F=2_1$--$1_0$) maps (not
repeated here) to gauge the velocity field of the L1148 region on very
large spatial scales (i.e., $\gg{}1\arcmin$, or $\gg{}0.1~\rm{}pc$).
The maps reveal similar velocities $\sim{}2.7~\rm{}km\,s^{-1}$ for the
two dust filaments (that are traced by the CCS intensity), and full
width at half maximum (FWHM) line widths (from Gaussian fits) of
$0.15~{\rm{}to}~0.25~\rm{}km\,s^{-1}$. There is a velocity
gradient along the north--western dust filament harboring L1148--IRS:
the filament's north--eastern edge (that contains L1148--IRS) has a
velocity $2.63~\rm{}km\,s^{-1}$, while the south-western end moves
with $2.78~\rm{}km\,s^{-1}$. This difference of $0.15~\rm{}km\,s^{-1}$
is similar to the line width, and thus rather significant. The
transition from ``low'' to ``high'' velocity occurs rather abruptly,
i.e., over a distance of only $\approx{}1\arcmin$, or
$\approx{}0.1~\rm{}pc$. This suggests that this filament actually
consists of two physically independent objects that form a single
object only in projection. The line width is slightly elevated in the
transition region, but not to a level which would clearly indicate the
physical interaction of the two velocity components (e.g., not by a
multiple of the average line width). The transition zone lies only
$\approx{}1\arcmin$ to the south--east of L1148--IRS, and is thus
relatively close to our target object.

\begin{figure}
  \begin{center}
    \includegraphics[height=\linewidth,angle=-90]{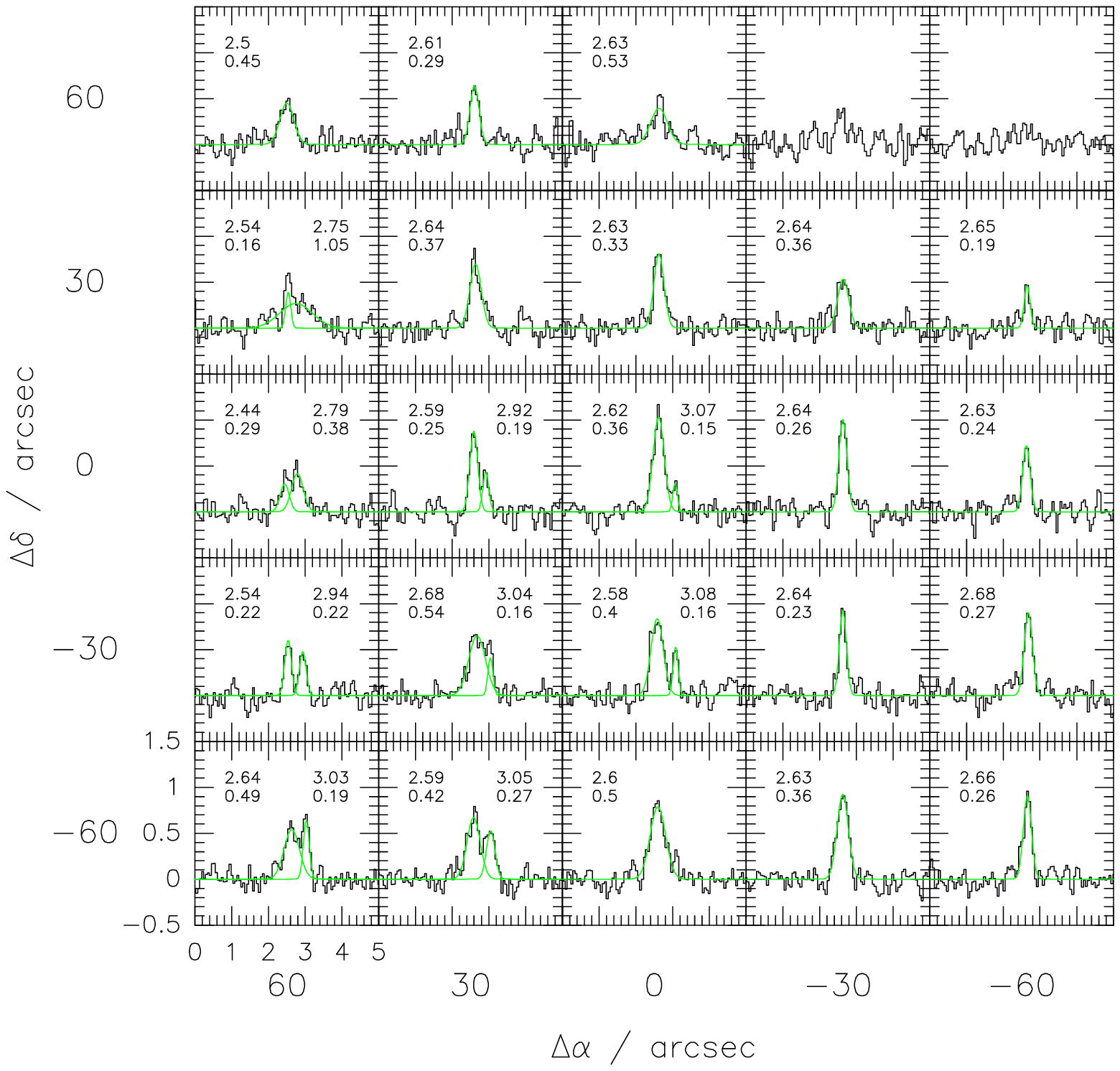}
    \caption{Spectra in the $\rm{}C^{18}O$ $J=2$--1 transition towards
      L1148--IRS as observed with frequency switching on a $5\times{}5$
      point grid with $30\arcsec$ spacing. Offsets are given relative
      to the approximate dust emission peak position
      ($\alpha=\rm{}20^h40^m56\fs{}5$,
      $\delta=+67\degr{}22\arcmin{}58\farcs{}0$ [J2000.0]). The LSR
      velocity and intensity in the $T_{\rm{}mb}$--scale are shown for
      the spectrum in the lower left corner. Gaussian fits to velocity
      components are drawn for those positions where the component
      identification is obvious. Small numbers give for every
      component fitted the derived LSR velocities (upper lines) and
      line width (lower lines; both in
      $\rm{}km~s^{-1}$).\label{app-fig:L1148_C18O_spec}}
  \end{center}
\end{figure}

In the maps of \citet{kauffmann2005:vellos}, smaller spatial scales
are well probed by the $\rm{}C^{18}O$ ($J=2$--1) maps of size
$100\arcsec{}\times{}100\arcsec$, or $(0.16~\rm{}pc)^2$. As shown in
Figure \ref{app-fig:L1148_C18O_spec}, they reveal two velocity
components in the immediate vicinity of L1148--IRS: one at velocities
$\sim{}2.6~\rm{}km\,s^{-1}$, and one at
$\sim{}3.0~\rm{}km\,s^{-1}$. The FWHM line widths (from Gaussian fits)
show significant variation across the map; they are
$0.2~{\rm{}to}~0.5~\rm{}km{}\,s^{-1}$, when only taking reliable fits
into account. The velocity of the lower--velocity emission globally
increases from north--east to the south--west, but this trend is not as
distinct as seen in CCS. Interestingly, the $\rm{}C^{18}O$ integrated
intensity presented in Figure \ref{app-fig:L1148_C18O_N2H+} does not
correlate with the dust intensity distribution (also see Appendix
\ref{app-sec:abundances}).

The kinematics of the very dense gas are traced by the $\rm{}N_2H^+$
($J=1$--0) maps of \citet{kauffmann2005:vellos}. As seen in Figure
\ref{app-fig:L1148_C18O_N2H+}, these cover the same region as the
$\rm{}C^{18}O$ maps. The spectra show no signs of multiple velocity
components. Characteristic velocities can be derived by simultaneously
fitting all $\rm{}N_2H^+$ hyperfine satellites. They are of order
$2.6~\rm{}km\,s^{-1}$. Velocities increase from north--east
($\sim{}2.55~\rm{}km\,s^{-1}$) to south--west
($\sim{}2.63~\rm{}km\,s^{-1}$), just as seen for CCS and
$\rm{}C^{18}O$. The FWHM line widths are
$0.15~{\rm{}to}~0.25~\rm{}km\,s^{-1}$, with a peak towards
L1148--IRS.\medskip

\begin{figure}
\includegraphics[height=\linewidth,angle=-90,bb=215 97 420 385,clip]{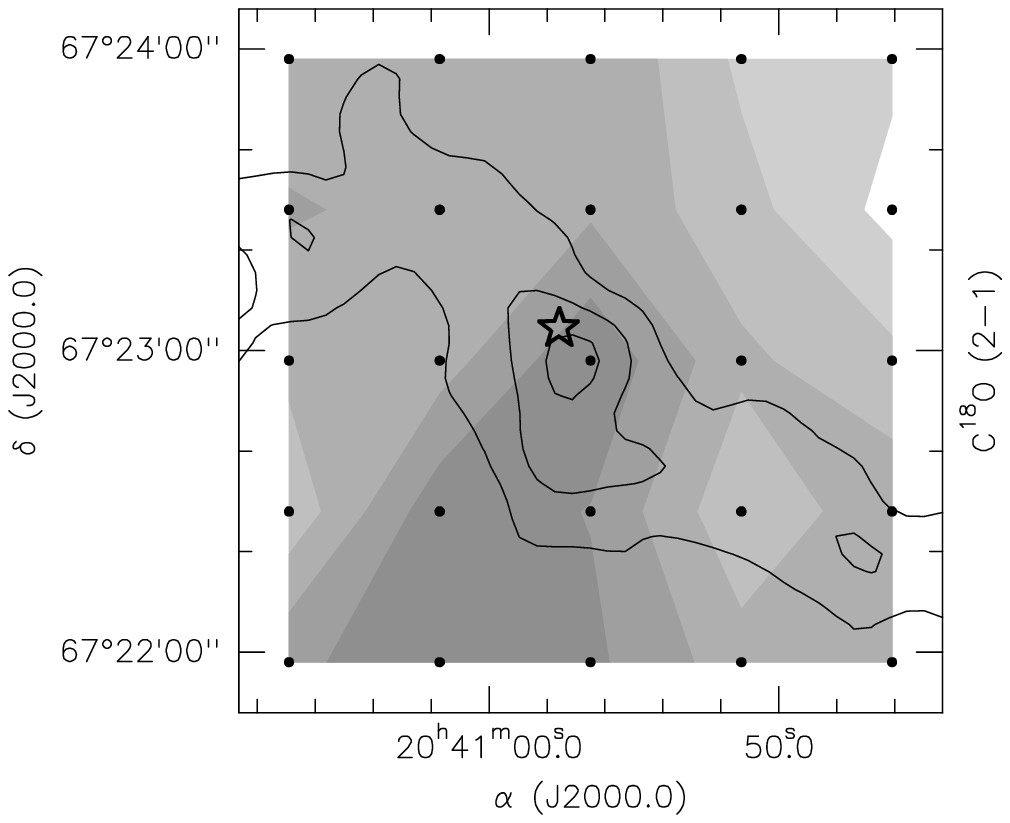}
\includegraphics[height=\linewidth,angle=-90,bb=216 97 450 385,clip]{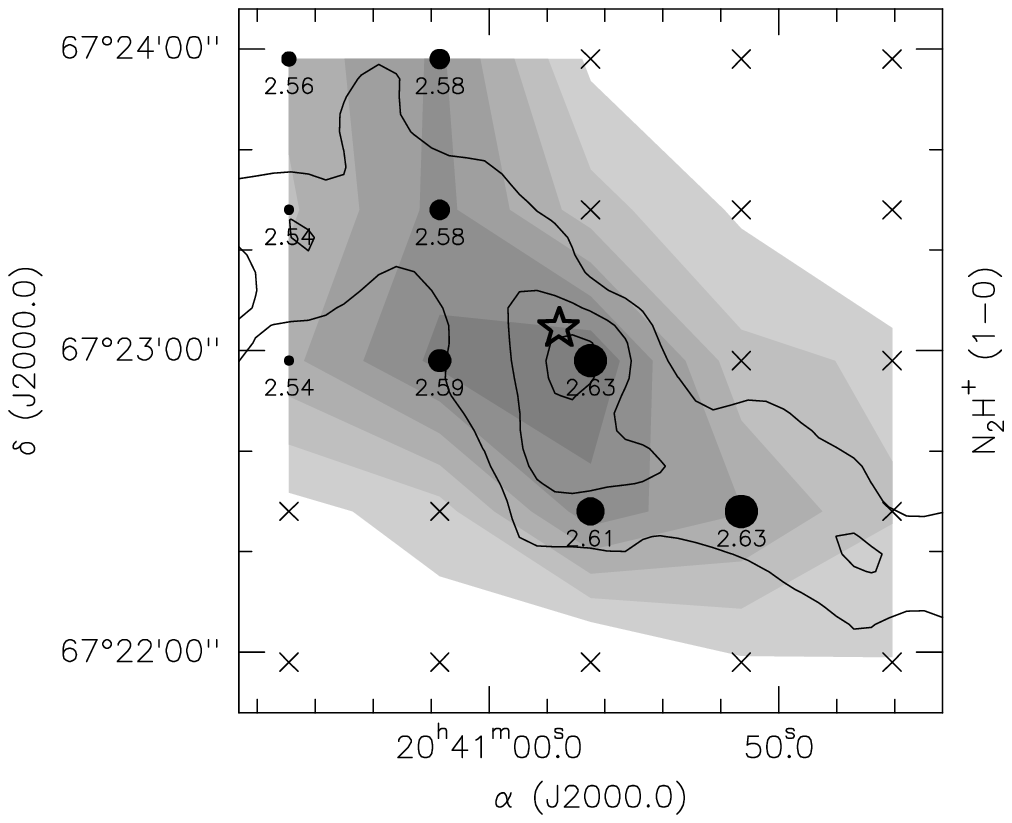}
\caption{Maps of $\rm{}C^{18}O$ ($J=2$--1) and $\rm{}N_2H^+$
  ($J=1$--0) integrated intensity. Grid markers give the positions
  observed in our study. Shading illustrates the intensities recorded
  during the observations. Contours outline the dust emission detected
  by MAMBO. For $\rm{}N_2H^+$, marker sizes and numbers indicate the
  radial velocity derived from hyperfine fitting; crosses are drawn if
  the emission is too faint to be fitted reliably. A star marks the
  position of L1148--IRS.\label{app-fig:L1148_C18O_N2H+}}
\end{figure}

\noindent{}The low observed line widths
$\lesssim{}0.25~\rm{}km\,s^{-1}$ are noteworthy; $\rm{}N_2H^+$ lines
in supposedly ``evolved'' cores \citep{crapsi2005:survey}, as well as
CCS transitions in star--forming cores
\citep{degregorio-onsalvo2006:ccs-in-yso}, are observed to typically
have lines wider than $0.25~\rm{}km\,s^{-1}$. It is, e.g., possible
that the wide lines in non--VeLLO cores are a manifestation of
large--scale rapid inward motions associated with accretion on forming
stars. If these motions are slower in VeLLO cores, or their spatial
extent is smaller, then if would be understandable why VeLLO cores
have so narrow lines. However, in the light of the present data
situation, any such consideration remains speculative.

\subsection{Molecular Abundances\label{app-sec:abundances}}
The CCS integrated intensity does not correlate well with the dust
emission (Figure 3 of \citealt{kauffmann2005:vellos}). 
In contrast, $\rm{}N_2H^+$ (1--0) emission (Figure
\ref{app-fig:L1148_C18O_N2H+}) traces the core very well.

The correlation between the $\rm{}C^{18}O$ (2--1) emission and the
dust distribution is unusually weak (Figure
\ref{app-fig:L1148_C18O_N2H+}): while the dust emission filament runs
NE-SW, the $\rm{}C^{18}O$ emission features a distinct wedge that
starts near L1148--IRS and flares towards SSE (position angle
  roughly $160\degr$). This structure is part
of the aforementioned lower--velocity $\rm{}C^{18}O$ component: it is
still seen if subtracting the emission from the higher--velocity
component from the map (by using the fits shown in Figure
\ref{app-fig:L1148_C18O_spec}).\medskip

\noindent{}\citet{kauffmann2005:vellos} calculate
abundances using the procedures and molecular constants laid out by
\citet{crapsi2005:survey}; for $\rm{}N_2H^+$, optical
depths and excitation temperatures are derived from
hyperfine--fits, while $\rm{}C^{18}O$ lines are assumed to be optically
thin and to have an excitation temperature of $10~\rm{}K$. Under these
assumptions, the CO depletion factor (which is supposed to increase
when CO freezes out onto dust),
$f_{\rm{}d}=[N({\rm{}C^{18}O})/N({\rm{}H_2})]_0 /
[N({\rm{}C^{18}O})/N({\rm{}H_2})]_{\rm{}obs}$, is calculated from the
observed (subscript `obs') and ``canonic'' (`0') dust and CO column
densities, $N$, as
\begin{equation}
f_{\rm d} = 
0.42 \,
\left(
\frac{
  S_{\rm 1.2 ~ \rm mm}^{\rm beam}
}{
  \rm mJy \, beam^{-1}
}
\right) \,
\left(
\frac{
  \int T_{\rm mb}({\rm C^{18}O}[2-1]) \, {\rm d} v
}{
  \rm K \, km \, s^{-1}
}
\right)^{-1} \, ,
\end{equation}
where $S_{\rm{}1.2~\rm{}mm}^{\rm{}beam}$ is the dust emission
intensity (per $11\arcsec$ beam), and
$\int{}T_{\rm{}mb}({\rm{}C^{18}O}[2-1])\,{\rm d}v$ is the
velocity-integrated $\rm{}C^{18}O$ intensity in
$T_{\rm{}mb}$--scale. Note that \citet{kauffmann2005:vellos} use --- in
deviation to all other parts of this paper --- dust emission
intensity--to--mass conversion factors from \citet{crapsi2005:survey}
instead of those suggested by
\citet{kauffmann2008:mambo-spitzer}. This scheme is adopted to make
our depletion factors more comparable to those of
\citeauthor{crapsi2005:survey}. If one were to use the
\citeauthor{kauffmann2008:mambo-spitzer} recommendations, $f_{\rm{}d}$
would decrease by a factor 2.54. Comparison shows that this approach
typically overestimates the \citeauthor{crapsi2005:survey}
(\citeyear{crapsi2005:survey}; their Table 6) estimates from (1--0)
line data by a factor of 1.3, probably because of different excitation
conditions for the (1--0) and (2--1) lines and differences in beam
matching (for the present data, the beam in $\rm{}C^{18}O$
observations is smaller than the one in the smoothed MAMBO maps used
to get $S_{\rm{}1.2~\rm{}mm}^{\rm{}beam}$).

The CO depletion factor peaks towards L1148--IRS, where
$f_{\rm{}d}=7.8\pm{}1.3$. Compared to other dense cores, this is
relatively low, in particular since this approach is likely to
overestimate the true value (see above). For the same position, an
$\rm{}N_2H^+$ column density of
$(3.1{}\pm{}0.6)\times{}10^{12}~\rm{}cm^{-2}$ is derived from deep new
integrations to probe contraction motions (Appendix
\ref{app-sec:contraction}). We do not detect significant $\rm{}N_2D^+$
emission. Assuming that the lines are optically thin, we derive
$\rm{}N_2D^+$ column densities $<0.5\times{}10^{12}~\rm{}cm^{-2}$
(using the intensity noise level times three).

\subsection{Contraction Motions\label{app-sec:contraction}}
\begin{figure}
  \setlength{\tabcolsep}{1.5pt}
  \begin{tabular}{ccc}
    \includegraphics[scale=0.6,angle=-90]{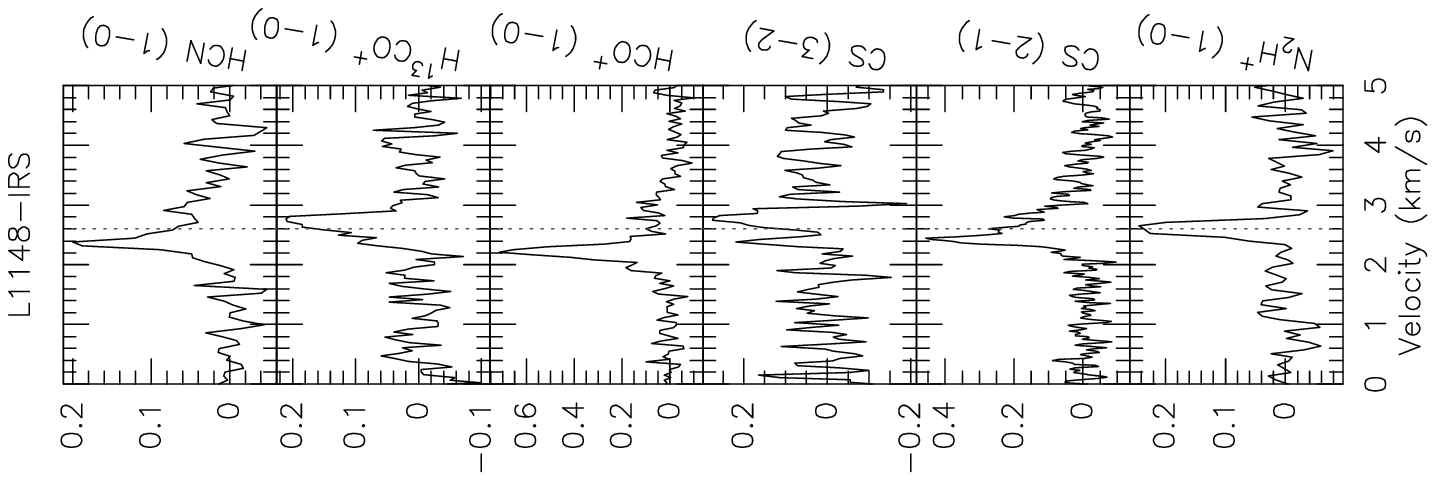}&
    \includegraphics[scale=0.6,angle=-90]{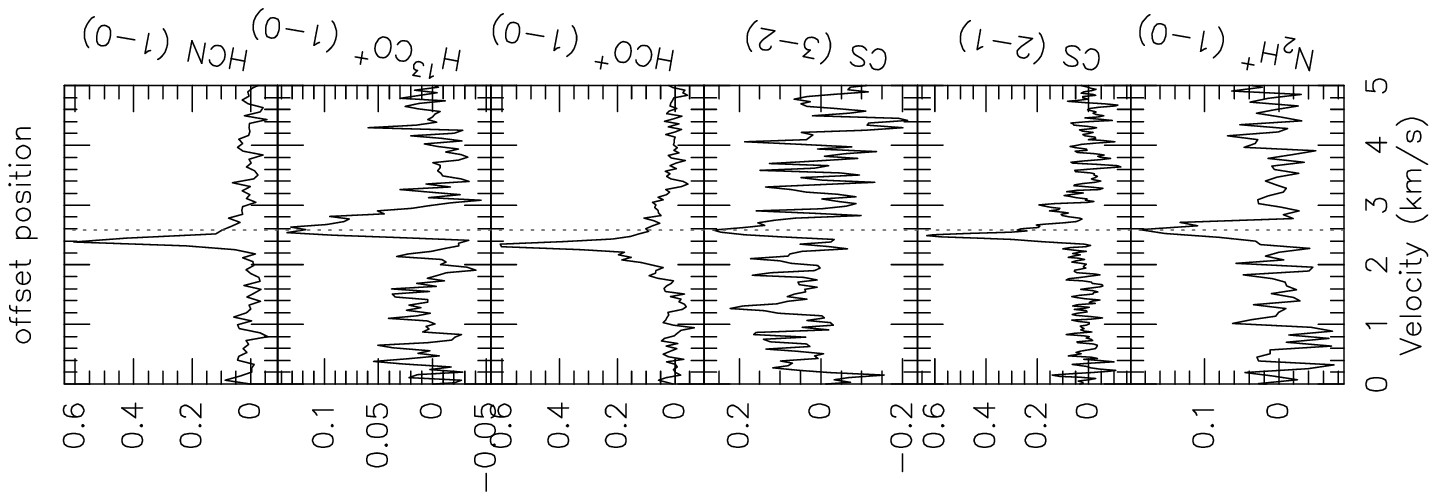}&
    \includegraphics[scale=0.6,angle=-90]{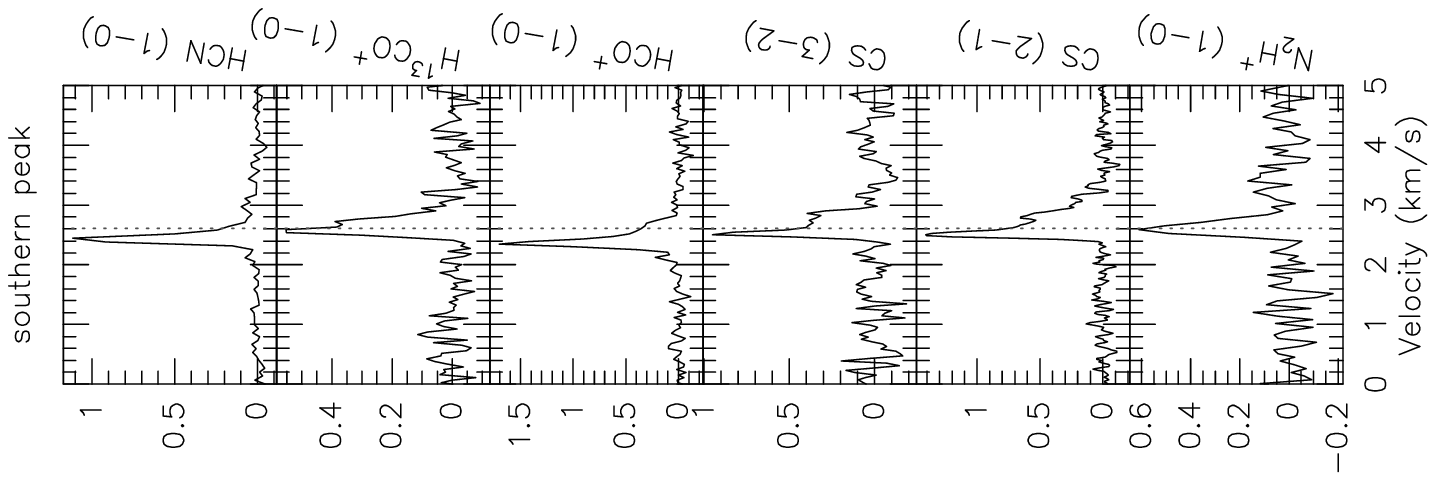}
  \end{tabular}
  \caption{Spectra towards the positions (left to right) and
    transitions (top to bottom; $J,F=1,2$--$0,1$ for HCN,
    $J,F_1,F=1,0,1$--$0,1,2$ for $\rm{}N_2H^+$)
    defined in Table \ref{app-tab:velshift_abs}. Dotted vertical lines
    indicate the systemic velocity derived from the $\rm{}N_2H^+$
    ($J,F_1,F=1,0,1$--$0,1,2$) transition. All well detected lines
    believed to be good tracers of infall have peaks blueshifted from
    the systemic velocity, indicating inward
    motions.\label{app-fig:velshift_all}}
\end{figure}

The systemic reference velocity of the dense core, $v_{\rm{}ref}$, can
be well measured using the optically thin
$J,F_1,F=1,0,1$--$0,1,2$ transition of the
$\rm{}N_2H^+$ dense gas tracer. Towards L1148--IRS, we derive a velocity
of $2.60~\rm{}km\,s^{-1}$ (Table \ref{app-tab:velshift_abs}). Many
supposedly optically thick lines do, however, peak at lower velocities
(Figure \ref{app-fig:velshift_all}), i.e.\ they
are blue--shifted. Such velocity shifts are characteristic of dense
cores undergoing contraction \citep{leung1977:inward_motions}. The
velocity offset of a line peaking at velocity $v_i$ can be
characterized using dimensionless velocity differences,
\begin{equation}
\delta v_i = \frac{v_i - v_{\rm ref}}{\Delta v_{\rm ref}}
\label{eq-app:velocity-differences}
\end{equation}
\citep{mardones1997:infall}, where $\Delta{}v$ is the line width at
half peak intensity. If $\delta{}v_i<0$, then the line is
blue--shifted, which suggests contraction motions. Expansion implies
$\delta{}v_i>0$.

\begin{table*}
\begin{minipage}{\linewidth}
  \caption{Line Velocities and Widths for several Transitions observed
    towards selected Positions.\label{app-tab:velshift_abs}}
    \begin{tabular}{lllllll}
      \hline
      Line & \multicolumn{2}{c}{L1148--IRS} &
      \multicolumn{2}{c}{offset position} &
      \multicolumn{2}{c}{southern peak}\\
       & \multicolumn{2}{c}{$\rm{}20^h40^m56\fs{}5$,
         $+67\degr{}22\arcmin{}58\farcs{}0$} &
      \multicolumn{2}{c}{$\rm{}20^h41^m01\fs{}7$,
        $+67\degr{}23\arcmin{}28\farcs{}0$} &
      \multicolumn{2}{c}{$\rm{}20^h41^m14\fs{}0$,
        $+67\degr{}20\arcmin{}31\farcs{}5$}\\
       & $v_{\rm LSR} / \rm km ~ s^{-1}$ & 
         $\Delta v / \rm km ~ s^{-1}$ &
         $v_{\rm LSR} / \rm km ~ s^{-1}$ & 
         $\Delta v / \rm km ~ s^{-1}$ & 
         $v_{\rm LSR} / \rm km ~ s^{-1}$ & 
         $\Delta v / \rm km ~ s^{-1}$\\ \hline
      $\rm N_2H^+$ ($J, F_1, F = 1, 0, 1$--$0, 1, 2$) &
        $2.602 \pm 0.015$ & $0.244 \pm 0.030$ &
        $2.583 \pm 0.023$ & $0.232 \pm 0.048$ &
        $2.614 \pm 0.013$ & $0.243 \pm 0.029$\\
      CS ($J = 2$--1) &
        $2.418 \pm 0.005$ & $0.149 \pm 0.017$ &
        $2.488 \pm 0.003$ & $0.118 \pm 0.010$ &
        $2.509 \pm 0.002$ & $0.119 \pm 0.004$\\
      CS ($J = 3$--2) &
        & &
        & &
        $2.513 \pm 0.005$ & $0.106 \pm 0.015$\\
      $\rm HCO^+$ ($J = 1$--0) &
        $2.218 \pm 0.001$ & $0.204 \pm 0.013$ &
        $2.327 \pm 0.003$ & $0.119 \pm 0.008$ &
        $2.356 \pm 0.002$ & $0.115 \pm 0.006$\\
      $\rm H^{13}CO^+$ ($J = 1$--0) &
        $2.681 \pm 0.016$ & $0.420 \pm 0.038$ &
        $2.652 \pm 0.015$ & $0.314 \pm 0.028$ &
        $2.628 \pm 0.008$ & $0.293 \pm 0.017$\\
      HCN ($J, F = 1, 2$--$0, 1$) &
        $2.360 \pm 0.013$ & $0.197 \pm 0.048$ &
        $2.398 \pm 0.006$ & $0.127 \pm 0.018$ &
        $2.424 \pm 0.008$ & $0.125 \pm 0.021$\\
      HCN ($J, F = 1, 1$--$0, 1$) &
        & &
        $2.429 \pm 0.018$ & $0.198 \pm 0.046$ &
        $2.438 \pm 0.012$ & $0.113 \pm 0.032$\\
      HCN ($J, F = 1, 0$--$0, 1$) &
        $2.601 \pm 0.041$ & $0.566 \pm 0.086$ &
        $2.460 \pm 0.016$ & $0.153 \pm 0.033$ &
        $2.469 \pm 0.010$ & $0.124 \pm 0.034$\\ \hline
    \end{tabular}

\vspace{0.5ex} \textbf{Notes.}
The observations hold for the J2000.0 positions defined in the
header. These are also marked in Figure \ref{fig:layout-observations}. If more
than one line has to be fitted simultaneously to achieve a good fit to
the narrow peak and the broad base of a line, then the parameters for the
component with the lower velocity are listed. No numbers are given for
transitions not detected well enough to allow fitting. The lines
believed to be optically thick all peak at velocities below the
systemic one derived from $\rm{}N_2H^+$.
\end{minipage}
\end{table*}

\begin{table}
  \caption{Dimensionless Velocity Shifts towards selected
    Positions\label{app-tab:velshift_rel}}
    \begin{tabular}{llll}
      \hline
      Line & \multicolumn{3}{c}{
        $\delta v_i = (v_i - v_{\rm N_2H^+}) / \Delta v_{\rm N_2H^+}$}\\
       & L1148--IRS & offset position & southern peak\\ \hline
       CS\\
       $J = 2$--1 & $-0.75 \pm 0.09$ & $-0.41 \pm 0.08$ &
         $-0.43 \pm 0.05$\\
       $J = 3$--2 & & & $-0.42 \pm 0.05$\\
       \\ $\rm HCO^+$\\
       $J = 1$--0 & $-1.57 \pm 0.19$ & $-1.10 \pm 0.23$ &
         $-1.06 \pm 0.13$\\
       \\ $\rm H^{13}CO^+$\\
       $J = 1$--0 & $0.32 \pm 0.04$ & $0.30 \pm 0.06$ &
         $0.06 \pm 0.01$\\
       \\ HCN\\
       $J, F = 1, 2$--$0, 1$ & $-0.99 \pm 0.12$ & $-0.80 \pm 0.16$ &
         $-0.78 \pm 0.09$\\
       $J, F = 1, 1$--$0, 1$ & & $-0.66 \pm 0.14$ &
         $-0.72 \pm 0.09$\\
       $J, F = 1, 0$--$0, 1$ & $0.00 \pm 0.10$ & $-0.53 \pm 0.11$ &
         $-0.60 \pm 0.07$\\ \hline
    \end{tabular}

\vspace{0.5ex} \textbf{Notes.}
Dimensionless velocity shifts in several transitions
towards the positions defined in Table \ref{app-tab:velshift_abs}. No numbers
are given for transitions not well enough detected to allow
fitting. The lines believed to be optically thick
all do have negative line shifts indicative of inward
motions.
\end{table}

Table \ref{app-tab:velshift_abs} gives observed peak velocities in
L1148. These are derived by fitting Gaussian lines to the peaks (and
sometimes additional Gaussians to the base). Table
\ref{app-tab:velshift_rel} gives the corresponding dimensionless
velocity differences. As outlined in Figure \ref{fig:layout-observations}, we
searched three positions for contraction motions, in order to remove
biases from possible foregrounds to L1148. All lines are supposedly
optically thick. This only excludes the ``isolated'' $\rm{}N_2H^+$
hyperfine satellite (i.e., $J,F_1,F=1,0,1$--$0,1,2$) and the
$\rm{}H^{13}CO^+$ (1--0) transition, which can therefore work as
velocity references.

All well--detected optically thick lines peak at velocities
blue--shifted from the $\rm{}N_2H^+$ reference velocity. This only
excludes the ($F=0$--1) emission of HCN towards L1148--IRS, which is a
particular case discussed below. In all tracers, the dimensionless
line asymmetries are largest towards L1148--IRS. This position is a
``strong infall candidate'', following the criteria for CS (2--1)
emission put forward by \citeauthor{lee1999:contr_survey}
(\citeyear{lee1999:contr_survey}: i.e., a CS emission peak with
significant blue--shift; red--shifted emission at a brightness below the
one of the blue--shifted emission; and no asymmetries in the isolated
$\rm{}N_2H^+$ line). Also, the values of the CS and HCN line
asymmetries towards L1148--IRS are unusually large for starless cores:
only 18\% of the CS cores probed by \citet{lee1999:contr_survey},
respectively 40\% of the HCN cores studied in the ($F = 2$--1)
transition by \citet{sohn2004:hcn}, have larger asymmetries (for CS,
we use the data after the reanalysis by
\citealt{lee2001:contr_survey}). Note that the asymmetry in the HCN
satellites decreases with decreasing optical depth (in LTE, the
optical depth ratios for different $F$ are 1:5:3 for
[0--1]:[2--1]:[1--1]), just as expected for contracting cores. Towards
L1148--IRS, the possibly optically thin $F=0$--1 line of HCN has an
unusually large line width of $0.57~\rm{}km\,s^{-1}$. This may hint at
mechanical energy input by the YSO, for example in the form of
outflows. The $\rm{}H^{13}CO^+$ (1--0) is supposedly optically thin
and peaks at higher velocities then the optically thick lines. This
confirms the velocity differences with respect to $\rm{}N_2H^+$,
although the line cannot serve as a precise velocity reference because
of unresolved hyperfine substructure \citep{schmid-burgk2004:l1512}.

It thus appears that the natal dense core of L1148--IRS is indeed
contracting. This is suggested by the negative velocity differences
and the particularly strong line asymmetries towards the embedded
source, L1148--IRS. The prevalence of blue--shifts throughout the L1148
region may actually indicate that the entire region (i.e., on scales
larger than the $\sim{}0.3~\rm{}pc$ containing the three positions
probed) is contracting.

\label{lastpage}

\end{document}